\documentclass[sn-mathphys-num]{sn-jnl}

\usepackage{lineno}
\usepackage{graphicx}%
\usepackage{multirow}%
\usepackage{amsmath,amssymb,amsfonts}%
\usepackage{amsthm}%
\usepackage{mathrsfs}%
\usepackage[title]{appendix}%
\usepackage[dvipsnames]{xcolor}%
\usepackage{textcomp}%
\usepackage{manyfoot}%
\usepackage{booktabs}%
\usepackage{algorithm}%
\usepackage{algorithmicx}%
\usepackage{algpseudocode}%
\usepackage{listings}%
\usepackage{units}
\usepackage{soul}

\theoremstyle{thmstyleone}%
\theoremstyle{thmstyletwo}%

\theoremstyle{thmstylethree}%

\newcommand{\pistroke}{
  \text{\protect\ooalign{\hidewidth\raisebox{-0.2ex}{--}\hidewidth\cr$\pi$\cr}}
}

\begin{document}
\title[Evidence of the pair instability gap from black hole masses]{Evidence of the pair instability gap from black hole masses}


\author*[1,2]{\fnm{Hui} \sur{Tong}}\email{hui.tong@monash.edu}
\author[3,4,5]{\fnm{Maya} \sur{Fishbach}}
\author[1,2]{\fnm{Eric} \sur{Thrane}}
\author[6,7,8,9]{\fnm{Matthew} \sur{Mould}}
\author[10,11]{\fnm{Thomas A.} \sur{Callister}}
\author[3,12]{\fnm{Amanda} \sur{Farah}}
\author[1,2]{\fnm{Nir} \sur{Guttman}}
\author[1,2]{\fnm{Sharan} \sur{Banagiri}}
\author[13]{\fnm{Daniel} \sur{Beltran-Martinez}}
\author[14]{\fnm{Ben} \sur{Farr}}
\author[15,16]{\fnm{Shanika} \sur{Galaudage}}
\author[14]{\fnm{Jaxen} \sur{Godfrey}}
\author[7,8,9]{\fnm{Jack} \sur{Heinzel}}
\author[17,18]{\fnm{Marios} \sur{Kalomenopoulos}}
\author[19,20]{\fnm{Simona J.} \sur{Miller}}
\author[3]{\fnm{Aditya} \sur{Vijaykumar}}

\affil*[1]{\orgdiv{School of Physics and Astronomy}, \orgname{Monash University}, \orgaddress{\city{Clayton}, \state{VIC} \postcode{3800}, \country{Australia}}}
\affil*[2]{\orgdiv{OzGrav: The ARC Centre of Excellence for Gravitational Wave Discovery}, \orgaddress{\city{Clayton}, \state{VIC} \postcode{3800}, \country{Australia}}}
\affil[3]{\orgdiv{Canadian Institute for Theoretical Astrophysics}, \orgname{University of Toronto}, \orgaddress{\street{60 St George St}, \city{Toronto}, \state{ON} \postcode{M5S 3H8}, \country{Canada}}}
\affil[4]{\orgdiv{David A. Dunlap Department of Astronomy and Astrophysics}, \orgname{University of Toronto}, \orgaddress{\street{50 St George St}, \city{Toronto}, \state{ON} \postcode{M5S 3H8}, \country{Canada}}}
\affil[5]{\orgdiv{Department of Physics}, \orgname{University of Toronto}, \orgaddress{\street{60 St George St}, \city{Toronto}, \state{ON} \postcode{M5S 3H8}, \country{Canada}}}
\affil[6]{\orgdiv{Nottingham Centre of Gravity \& School of Mathematical Sciences}, \orgname{University of Nottingham}, \orgaddress{\street{University Park}, \city{Nottingham}, \postcode{NG7 2RD}, \country{United Kingdom}}}
\affil[7]{\orgdiv{LIGO Laboratory}, \orgname{Massachusetts Institute of Technology}, \orgaddress{\city{Cambridge}, \state{MA} \postcode{02139}, \country{USA}}}
\affil[8]{\orgdiv{Kavli Institute for Astrophysics and Space Research}, \orgname{Massachusetts Institute of Technology}, \orgaddress{\city{Cambridge}, \state{MA} \postcode{02139}, \country{USA}}}
\affil[9]{\orgdiv{Department of Physics}, \orgname{Massachusetts Institute of Technology}, \orgaddress{\city{Cambridge}, \state{MA} \postcode{02139}, \country{USA}}}
\affil[10]{\orgname{Williams College}, \orgaddress{\city{Williamstown}, \state{MA} \postcode{01267}, \country{USA}}}
\affil[11]{\orgdiv{Kavli Institute for Cosmological Physics}, \orgname{The University of Chicago}, \orgaddress{\city{Chicago}, \state{IL} \postcode{60637}, \country{USA}}}
\affil[12]{\orgdiv{Department of Physics}, \orgname{The University of Chicago}, \orgaddress{\city{Chicago}, \state{IL} \postcode{60637}, \country{USA}}}
\affil[13]{\orgdiv{Centro de Investigaciones Energ\'eticas Medioambientales y Tecnol\'ogicas}, \orgaddress{\street{Avda. Complutense 40}, \city{Madrid}, \postcode{28040}, \country{Spain}}}
\affil[14]{\orgdiv{Institute  for  Fundamental  Science, Department of Physics}, \orgname{University of Oregon}, \orgaddress{\city{Eugene}, \state{OR} \postcode{97403}, \country{USA}}}
\affil[15]{
  \orgdiv{Laboratoire Lagrange}, 
  \orgname{Universit\'e C\^ote d'Azur, Observatoire de la C\^ote d'Azur, CNRS}, 
  \orgaddress{\street{Bd de l'Observatoire}, \postcode{06300}, \country{France}}
}
\affil[16]{
  \orgdiv{Laboratoire Artemis}, 
  \orgname{Universit\'e C\^ote d'Azur, Observatoire de la C\^ote d'Azur, CNRS}, 
  \orgaddress{\street{Bd de l'Observatoire}, \postcode{06300}, \country{France}}
}
\affil[17]{\orgdiv{Nevada Center for Astrophysics}, \orgname{University of Nevada}, \orgaddress{\city{Las Vegas}, \state{NV} \postcode{89154}, \country{USA}}}
\affil[18]{\orgdiv{Department of Physics and Astronomy}, \orgname{University of Nevada}, \orgaddress{\city{Las Vegas}, \state{NV} \postcode{89154}, \country{USA}}}
\affil[19]{\orgdiv{Department of Physics}, \orgname{California Institute of Technology}, \orgaddress{\city{Pasadena}, \state{CA} \postcode{91125}, \country{USA}}}
\affil[20]{\orgdiv{LIGO Laboratory}, \orgname{California Institute of Technology}, \orgaddress{\city{Pasadena}, \state{CA} \postcode{91125}, \country{USA}}}

\abstract{Stellar theory predicts a forbidden range of black-hole masses between ${\sim}50$--$130\,M_\odot$ due to pair-instability supernovae \citep{Fowler:1964zz, 1967ApJ...148..803R, 1967_PhysRevLett.18.379, 1968Ap&SS...2...96F, Heger:2001cd, Woosley:2007qp, Farmer_2019}, but evidence for such a gap in the mass distribution from gravitational-wave astronomy has proved elusive.
Early hints of a cutoff in black-hole masses at ${\sim} 45\,M_\odot$ disappeared with the subsequent discovery of more massive binary black holes \citep{2019ApJ...882L..24A, gwtc2_pop}.
Here, we report evidence of the pair-instability gap in LIGO--Virgo--KAGRA's fourth gravitational wave transient catalog (GWTC-4), with a lower boundary of $44_{-4}^{+5} M_\odot$ (90\% credibility). 
While the gap is not present in the distribution of \textit{primary} masses $m_1$ (the bigger of the two black holes in a binary system), it appears unambiguously in the distribution of \textit{secondary} masses $m_2$, where $m_2 \leq m_1$.
The location of the gap lines up well with a previously identified transition in the binary black-hole spin distribution; binaries with primary components in the gap tend to spin more rapidly than those below the gap. 
We interpret these findings as evidence for a subpopulation of hierarchical mergers: binaries where the primary component is the product of a previous black-hole merger and thus populates the gap.
Our measurement of the location of the pair-instability gap constrains the $S$-factor for $^{12}\rm{C}(\alpha,\gamma)^{16}\rm{O}$ at 300keV to $260_{-108}^{+190}$ keV barns. 
}

\maketitle

Stellar theory predicts a lack of black holes with masses from ${\sim}50\,M_\odot$ to ${\sim}130\,M_\odot$ \citep{Fowler:1964zz, 1967ApJ...148..803R, 1967_PhysRevLett.18.379, 1968Ap&SS...2...96F, Heger:2001cd, Woosley:2007qp, Farmer_2019}. 
Stars with initial (zero-age main sequence) masses in the range 100 to 260 $M_\odot$ are expected to experience (pulsational) pair-instability supernovae. At these masses, the carbon--oxygen stellar core is sufficiently hot that photons spontaneously produce electron--positron pairs. This leads to a drop in photon pressure that triggers sudden gravitational collapse, followed by the explosive ignition of oxygen. The resulting stellar explosion is so powerful that it can disrupt the entire stellar core in a pair-instability supernova, leaving behind no remnant whatsoever. This manifests as a gap in the black hole mass spectrum in the $50$--$130\,M_\odot$ range. Alternatively, it may trigger a series of pulses that shed enough mass so that the star collapses to a black hole below the lower edge of the gap.

Although the existence of pair instability supernovae is a robust prediction of stellar theory, observational evidence has proved elusive. 
Few examples of (pulsational) pair instability supernova have been observed; promising candidates are the supernova SN2018ibb and SN2020acct~\citep{Schulze:2023vik, angus}.
Gravitational-wave observatories provide a new, promising probe of pair-instability supernovae because they are sensitive to black holes in exactly the relevant mass range~\citep{aLIGO,aVIRGO,KAGRA:2020tym}. 
Measurements of the mass gap can constrain important theoretical uncertainties, including the role of metallicity, neutrino physics, convective mixing efficiency, and nuclear reaction rates~\citep{Farmer_2019}.
For example, a driving uncertainty in the gap's location is the uncertain $^{12}\rm{C}(\alpha,\gamma)^{16}\rm{O}$ reaction rate, opening up the possibility to measure this with gravitational-wave observations~\cite{Farmer_2019,Farmer:2020xne}.

Previous gravitational-wave analyses have not found a clear gap in the distribution of black hole masses~\citep{gwtc3_pop, Edelman:2021fik} (though see Refs.~\cite{Mould:2022ccw, Wang:2022gnx, Li_2023yyt, antonini2024} which found hints that black holes above ${\sim}40\,M_\odot$ do not originate from stellar collapse using LIGO--Virgo--KAGRA's third gravitational wave transient catalog (GWTC-3)).
Gravitational waves encode the primary mass $m_1$ and secondary mass $m_2<m_1$ of each binary black hole. 
The first gravitational-wave transient catalog (GWTC-1) hinted at a cutoff in the primary mass distribution at ${\sim}45\,M_\odot$~\citep[e.g.,][]{2019ApJ...882L..24A}, but subsequent discoveries of more massive black holes ruled out a sharp $m_1$ cutoff~\citep{gwtc2_pop}.
Meanwhile, an observed ${\sim}35\,M_\odot$ peak in the distribution of black-hole primary mass was hypothesized to result from a pileup from pulsational pair-instability supernovae \cite{mass,gwtc2_pop}, but subsequent work suggests this interpretation is in tension with stellar physics, observed supernovae rates and nuclear physics \cite{Golomb:2023vxm,Stevenson:2019rcw}, though see, e.g., Ref.~\cite{Croon} for an alternative view.

While the distribution of primary black-hole masses does not show a clear gap, there have not been studies looking for a gap in the secondary-mass distribution.
Black holes that do not form directly from ordinary stellar collapse---such as those that originate from previous black-hole mergers~\citep{GerosaFishbach}, stellar mergers~\citep{DiCarlo:2019pmf, Renzo:2020smh}, collapsars with significant mass loss from above the gap~\citep{Siegel:2021ptt}, or which have experienced significant accretion~\citep{McKernan_2012}---may contaminate the gap. Such processes are expected to be rare, and therefore unlikely to affect both black holes in a binary.
We therefore analyze data from GWTC-4~\citep{GWTC-4_result} using a model that specifically searches for a mass gap in the distribution of $m_2$. 
Our model allows for an interval corresponding to the pair-instability gap in $m_2$, where the probability density is zero \citep{Edelman:2021fik}; see Equation~\eqref{eq:gap_model}. The absence of a gap, i.e., the width of the gap to be $<20\,M_\odot$, is ruled out at 99.9\% credibility (see Extended Data Figure~\ref{fig:corner_m2gap}).
The gap model is preferred over the no-gap model by a natural logarithm Bayes factor $\ln {\cal{B}}$ of ${\sim} 4.4$; the priors on the gap parameters as well as the details of the mass and spin models are described in Section~\ref{sec:mass_model}.

\begin{figure}
    \centering
    \includegraphics[width=1.0\linewidth]{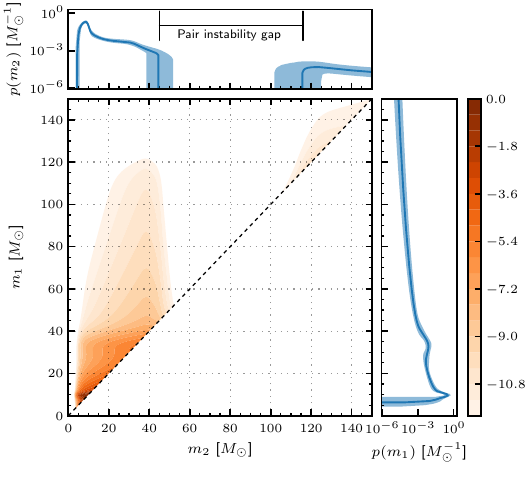}
    \caption{Reconstructed distribution of binary black-hole masses. The primary mass $m_1$ is by definition larger than the secondary mass $m_2 < m_1$.
    The mean prediction of the joint distribution $\pi(m_1, m_2)$ is shown in orange while the median prediction of the marginal distributions $\pi(m_1)$ and $\pi(m_2)$ are shown in solid blue with 90\% credibility range indicated by the shaded bands.
    The color bar represents the two-dimensional probability density $\log[p(m_1,m_2)/\max{p(m_1,m_2)}]$.
    The ``island'' of probability at the upper right of the two-dimensional density plane is mostly associated with the high-mass event GW231123.
}
    \label{fig:reconstructed_2D_mass_PPD}
\end{figure}

Figure~\ref{fig:reconstructed_2D_mass_PPD} shows the black hole mass distribution inferred under this model. 
While the $m_1$ distribution extends unbroken from $45 M_\odot$ to ${\sim}120M_\odot$, there is a clear gap in the $m_2$ distribution with the lower edge at $45_{-6}^{+7} M_\odot$ (90\% credibility), remarkably consistent with theoretical predictions for the pair-instability gap \citep{Fowler:1964zz, Woosley:2016hmi}. 
We constrain the upper edge of the gap to be $116_{-14}^{+9}M_\odot$ (90\% credibility). 
However, the upper edge is primarily constrained by GW231123, the most massive binary detected so far \citep{GW231123}.
The the upper edge uncertainty is dominated by uncertainty in the parameters of this single event, which may or may not originate from stars massive enough to collapse to the ``far side" of the gap, potentially experiencing significant mass loss via disk winds and jet launching~\citep{GW231123, Gottlieb:2025ugy}.
In contrast, the lower edge of the gap is a robust feature of the binary black hole population: excluding GW231123 from the analysis results in a lower edge of $43_{-6}^{+7}\,M_\odot$ and leaves the upper edge unconstrained. 
Such a gap is also present in the maximum population likelihood distribution~\citep{PhysRevResearch.5.023013} (see Extended Data Figure~\ref{fig:m1m2_pi_stroke} and the discussion in Section.~\ref{sec:pi_stroke}). 

The presence of the pair instability gap in the distribution of the secondary mass $m_2$ rather than the primary mass $m_1$ can be understood in the context of hierarchical mergers. Hierarchical mergers include one or more ``second-generation'' (2G) remnants of previous black-hole mergers.
Unlike ``first-generation" (1G) black holes that are born from stellar collapse, 2G black holes can populate the pair-instability gap.
Hierarchical mergers can occur in dense stellar environments~\citep{GerosaFishbach}.
For hierarchical mergers in dense star clusters with escape speeds typical of globular clusters, most 2G black holes are ejected from the cluster due to the recoil kicks they receive from the anisotropic emission of gravitational waves~\citep{Fitchett_1983,Gerosa:2018qay}.
Only a small fraction are retained that can merge again with another black hole~\citep{Antonini:2016gqe, Rodriguez:2019huv}. 
The merger of \textit{two} second-generation black holes (2G+2G) is expected to be comparatively rare. For example, the fraction of second-generation and first-generation black hole mergers (2G+1G) in globular clusters can be more than one order of magnitude higher than 2G+2G mergers \citep{Rodriguez:2019huv}. Taking selection effects into account, 2G+2G mergers could constitute at most 1\% of all detected binary black hole mergers if black holes are born with non-negligible spin~\citep{Rodriguez:2019huv}, as hinted by other analyses \citep{gwtc3_pop, GWTC_4_population}. The gap that we identify in the $m_2$ distribution may therefore represent the dearth of 2G+2G mergers. The prediction of 2G+2G rates in AGN is much more uncertain, replying on the details of gas-assisted migration.

Second-generation black holes are expected to merge with significant spins, inherited from the orbital angular momentum of their progenitor binaries.
If they are indeed hierarchical mergers, we therefore expect the binaries with $m_1 \gtrsim 45 M_\odot$ to include a primary with dimensionless spin of $\chi_1 \approx 0.7$.
Thus, measurements of black hole spin provide a means to independently verify our interpretation of a pair-instability mass gap contaminated with hierarchical mergers. Previous studies with GWTC-3 ~\citep{Li_2023yyt, Pierra_2024, antonini2024} have shown a transition of spin properties of mergers; Ref.~\citep{Li_2023yyt} first measured the transition at around the mass of $47^{+46}_{-10}\,M_\odot$ and Ref.~\citep{antonini2025_PISN} found this transition at $m_1=45^{+7}_{-5}\,M_\odot$ and associated that with the pair-instability gap. The spin distribution of the massive binary black holes is consistent with the prediction of hierarchical mergers.
To this end, we reanalyse GWTC-4 using the mass-dependent spin model developed in Ref.~\cite{antonini2024}.
In this model, the distribution of binary black-hole effective inspiral spin $\chi_{\rm{eff}}$, a weighted sum of the black hole spin from each component black hole, differs depending on whether the primary mass is below or above some transition mass $\tilde{m}_1$
(See Section~\ref{sec:spin_model} for additional details).
Given our astrophysical interpretation of the $m_2$ mass gap, we expect the spin transition mass $\tilde{m}_1$ to be consistent with the lower edge of the mass gap visible in $m_2$.

\begin{figure}
    \centering
    \includegraphics[width=1.0\linewidth]{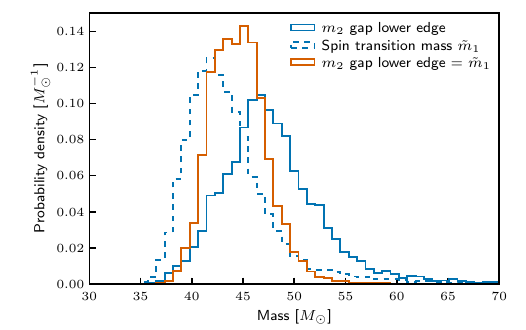}
    \caption{
    Posterior for the lower edge of the mass gap as determined by different properties.
    We plot the posterior on the spin transition mass $\tilde{m}_1$ with dashed line in blue; see Equation~\eqref{model:spin_simple} for the definition.
    We plot the posterior on the lower edge of the gap in $m_2$ with solid line in blue. These two mass scales are independently but simultaneously inferred using the model allowing both $m_2$ gap and spin transition mass $\tilde{m}_1$.
    In orange, we plot the posterior on the mass scale when the lower edge of the $m_2$ mass gap is taken to be the same as $\tilde{m}_1$.
    }
    \label{fig:comparison_spin_transition_mass}
\end{figure}

The results are summarized in Figure~\ref{fig:comparison_spin_transition_mass}. 
We plot the posterior on the spin transition mass $\tilde{m}_1$ with dashed blue.
It is strikingly consistent with the posterior on the lower edge of the $m_2$ mass gap, which is plotted with solid blue.
We conclude that the distribution of black-hole masses and the distribution of black-hole spins provide concordant evidence for a pair-instability gap contaminated by 2G+1G hierarchical mergers.
If we force the lower edge of the $m_2$ mass gap to be the same as $\tilde{m}_1$, we obtain the orange posterior, which implies a lower edge of $44_{-4}^{+5}\,M_\odot$. This model, in which the spin transition mass is fixed to the lower edge of the $m_2$ gap, is favoured over the transition-spin model without an $m_2$ gap by a $\ln {\cal{B}}$ of ${\sim}3.3$, and we adopt it for the remaining analyses unless otherwise specified. 

\begin{figure}[htp]
    \centering
    \includegraphics[width=1.0\linewidth]{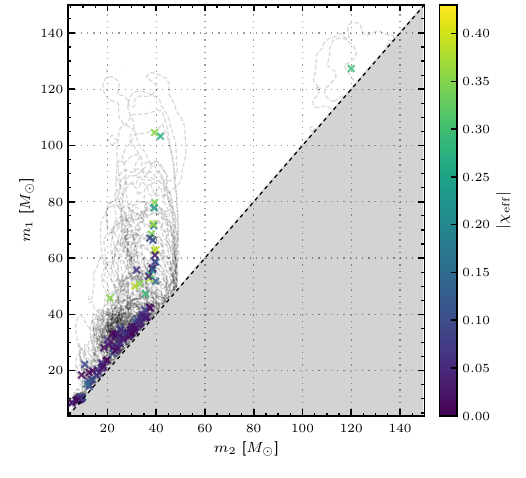}
    \caption{Primary masses $m_1$ and secondary masses $m_2$ coloured according to the absolute magnitude of $\chi_{\mathrm{eff}}$ for GWTC-4 events. The masses and spins of each event are simultaneously inferred with the population. We fit the population to a model that allows for a gap in $m_2$ and a $\chi_\mathrm{eff}$ spin distribution that transitions at the lower edge of the $m_2$ gap. Crosses show the median values of $m_1$ and $m_2$. Contours show 90\% credible intervals.
    }
    \label{fig:population_reweighted_scatter}
\end{figure}

Using the inferred population distribution as a prior to inform the masses of individual events, Figure~\ref{fig:population_reweighted_scatter} shows the population-informed masses $m_1$ and $m_2$ for events in GWTC-4. Each event is colored by the posterior median of the absolute magnitude of its inspiral spin parameter $|\chi_{\rm{eff}}|$. We can see a clear gap in $m_2$ among the observed events, together with the transition from small to large $\left|\chi_\text{eff}\right|$ happening around the same mass scale in $m_1$ as the lower edge of the gap in $m_2$. 
While hierarchical mergers in dense stellar clusters are usually predicted to have isotropic spin orientations due to dynamical assembly, preferentially aligned spins can arise from black holes in AGN disks or interactions between black holes and stars in clusters~\citep[e.g.,][]{McKernan:2023xio}. 
In Section~\ref{sec:spin_symmetry}, we test the symmetry of the spin distribution for the high mass mergers and show it is consistent with being symmetric around zero, matching the prediction of 2G+1G mergers from dense stellar clusters.

By assuming all mergers with $m_1$ inside the gap are 2G+1G hierarchical mergers, we infer a lower limit on the 2G+1G merger rate of $2.5_{-1.3}^{+2.1}\times10^{-1} \rm{Gpc}^{-3}\,\rm{yr}^{-1}$. We also put an upper limit on the rate of mergers with both components in the gap $< 7.9\times 10^{-2}\,\rm{Gpc}^{-3}\,\rm{yr}^{-1}$ at 90\% credibility (see Section~\ref{sec:notch_filter} for details).
By checking for events with primary masses unambiguously in the mass gap and secondary masses below the gap, we identify individual events that are likely to be 2G+1G hierarchical mergers.
Using the population-informed masses, we calculate the Bayes factor ${\cal B}$ between the hypothesis that an event contains a black hole with primary mass inside the gap to the hypothesis that it does not.
We find four events with strong support for the hierarchical-merger hypothesis ($\ln {\cal{B}} > 8$): GW190519\_153544, GW190602\_175927, GW191109\_010717 and GW231102\_071736.
The exceptional event GW190521 (with total mass $\sim 150 M_\odot$) is conspicuously absent from this list. 
This is because it is consistent with the ``straddling binary'' hypothesis \citep{Fishbach_GW190521}, where the primary mass exceeds the upper edge of the mass gap and the secondary mass falls below the lower edge, although it still has a $\ln {\cal{B}}$ of 1.7 in favour of the primary component in the gap (see Section~\ref{sec:GW190521} for additional details).
We estimate the merger rate of straddling binaries with one component from the far side of the gap and the other component below the gap to be $5.8^{+16.6}_{-4.6}\times 10^{-3}\,\rm{Gpc}^{-3}\,\rm{yr}^{-1}$. Meanwhile, the rate of mergers with both components on the far side of the gap is $3.8^{+4.6}_{-2.2}\times 10^{-2}\,\rm{Gpc}^{-3}\,\rm{yr}^{-1}$. Assuming a GWTC-4 cumulative search sensitivity, we expect $53^{+25}_{-20}$ in 300 binary black hole events to have primary masses in the gap and secondary masses below the gap. Meanwhile, we infer $6^{+7}_{-5}$ in 300 binary black hole events to contain at least one component on the far side of the gap, out of which $0^{+3}_{-0}$ can be straddling binaries. Under our best model, which indicates no detections of binaries with both components in the mass gap, Poisson uncertainty implies a 90\% upper limit of $\lesssim 4.5$ such systems in a catalog of 300 binary black hole detections.

\begin{figure}
    \centering
    \includegraphics[width=1.0\linewidth]{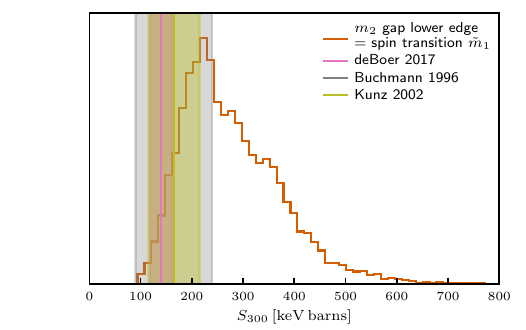}
    \caption{
    Constraints on $^{12}\rm{C}(\alpha,\gamma)^{16}\rm{O}$ rate.
    Orange is the posterior using the measurement imposing lower edge of the $m_2$ gap equal to the spin transition mass $\tilde{m}_1$.
    Pink shows the theoretical nuclear physics prediction from Ref.~\cite{deBoer_2017} of $\unit[140_{-21}^{+21}]{keV \, barns}$ (68\% credibility). Grey shows the result from Ref.~\cite{Buchmann_1996} of $\unit[165_{-75}^{+75}]{keV \, barns}$ (68\% credibility) and yellow-green shows the result from Ref.~\citep{Kunz_2002} of $\unit[165_{-50}^{+50}]{keV \, barns}$ (68\% credibility).}
    \label{fig:nuclear physics}
\end{figure}

The location of the pair-instability gap measurement can be used to constrain the $^{12}\rm{C}(\alpha,\gamma)^{16}\rm{O}$ rate~\citep{Farmer_2019, Farmer:2020xne, Golomb:2023vxm}. In astrophysics, the reaction cross section is commonly described by the $S$-factor, which is the nuclear structure-dependent part of the cross section for charged-particle nuclear reactions––isolated from Coulomb repulsion. 
Using the relationship between the lower edge of the mass gap and the $^{12}\rm{C}(\alpha,\gamma)^{16}\rm{O}$ rate fit from Ref.~\cite{Farmer:2020xne}, we constrain the astrophysical $S$-factor of $^{12}\rm{C}(\alpha,\gamma)^{16}\rm{O}$ at $\unit[300]{keV}$ to 
$\unit[260_{-108}^{+190}]{keV \, barns}$ (90\% credibility), as seen in Figure~\ref{fig:nuclear physics}. Our constraint on the $S$-factor is consistent with that of Ref.~\citep{deBoer_2017} with the median prediction of Ref.~\citep{deBoer_2017} within the 93\% credible interval, though our posterior favours higher values, which is more consistent with the values reported, e.g., in Refs.~\citep{Buchmann_1996, Kunz_2002, 2023ApJ...945...41S}. However, we do not interpret our result to be in tension with Ref.~\citep{deBoer_2017}; see Sec.~\ref{sec:S-factor} for more discussion. 

Going forward, it is important to include the $m_2$ gap in binary black hole population models.
The gap can potentially be used to break the degeneracy between distance and redshift, facilitating measurements of the Hubble parameter \citep{Farr_2019}. 
While we interpret the dearth of binary black holes with $m_2 \gtrsim 45 M_\odot$ as evidence of a pair instability gap---and while we regard this explanation as a promising and plausible explanation---we acknowledge that the data are open to other interpretations. For example, one could imagine that a dearth of events with $m_2 \gtrsim 45 M_\odot$ might be produced through the complicated interplay between the initial stellar mass function and binary interactions, thereby mimicking a pair instability gap through unrelated physical processes. If subsequent data continue to show a gap for binaries with $m_2 \gtrsim 45 M_\odot$ regardless of the mass of the primary black hole, alternative explanations of the gap invoking binary interactions may begin to appear contrived. More excitingly, future measurements of spin orientations may help indicate the evolutionary pathways and formation environments of high-spin events to further test the hierarchical merger hypothesis.

\section{Methods}\label{sec:method}

\renewcommand{\figurename}{Extended Data Figure}
\renewcommand{\tablename}{Extended Data Table}

\subsection{Hierarchical Bayesian inference details}

We perform hierarhical Bayesian inference with \texttt{GWPopulation} \cite{gwpopulation} on the recently released cumulative Gravitational Wave Transient Catalog 4 \citep{Fishbach:2018edt, 2019_Bayesian, Mandel_2019, Vitale:2020aaz, GWTC-4_result, GWTC-4_method}. 
There are roughly 80 confident binary black holes detected in the first part of the LIGO--Virgo--KAGRA's fourth observing run (O4a), enabled by a variety of detector improvements \citep{2020PhRvD.102f2003B, 2023PhRvX..13d1021G, LIGOScientific:2024elc, Capote:2024rmo, LIGO:2024kkz}.
Our dataset includes 153 binary black holes with false-alarm rates $\le 1\,\rm{yr}^{-1}$, consistent with Ref.~\cite{GWTC_4_population}.
Selection effects are taken into account by a Monte Carlo estimation \cite{Tiwari:2017ndi, 2019_Farr_selection, Essick:2022ojx, Essick:2025zed}. 
We use the posterior samples consistent with the choice of Ref.~\cite{GWTC_4_population}, generated using \texttt{Bilby} and \texttt{Dynesty} \cite{2019_bilby, 2020_bilby, dynesty}.
For events detected before O4a, we use the MIXED samples reported in GWTC-3 and GWTC-2.1 \citep{gwtc-2.1,gwtc-3}. 

As the previously known event with the highest support for pair-instability gap components GW190521, Ref.\citep{GW190521_implications} concluded the \textsc{NRSur7dq4} waveform model is most faithful
to NR simulations in the parameter range relevant for
this exceptional event. 
We tested different waveform models for this event and concluded the results of our pair-instability gap measurement are not sensitive to the choice of waveform.
Using \textsc{NRSur7dq4} posterior samples of GW19021, we found the lower edge of $49^{+10}_{-6}$ from mass distribution and $45^{+5}_{-3}$ by incorporating spin transition information.

\subsection{Basic mass gap model}\label{sec:mass_model}

To model the predicted pair-instability gap~\citep{Fowler:1964zz, 1967ApJ...148..803R, 1967_PhysRevLett.18.379, 1968Ap&SS...2...96F, Heger:2001cd, Woosley:2007qp, Woosley:2016hmi, Woosley_2019,Farmer_2019, Marchant:2018kun, Woosley:2021xba, Renzo:2024jhc, Spera:2017fyx}, we employ an extension of the basic binary black-hole mass models in Ref.~\cite{GWTC_4_population}. 
The mass model is parameterized via source-frame primary mass $m_1$ and secondary mass $m_2<m_1$, so we are fitting the distribution 
\begin{align}
    \pi(m_1, m_2 | \Lambda) = \pi(m_1 | \Lambda) \, \pi(m_2 | m_1, \Lambda),
\end{align}
which is conditioned on hyper-parameters $\Lambda$ that control the shape of the distribution.
As discussed below, we employ various models for $\pi(m_1|\Lambda)$ from Ref.~\cite{GWTC_4_population}.
The key ingredient, however, is our model for secondary mass, which enforces a gap in the distribution of $m_2$:
\begin{align}\label{eq:gap_model}
    \pi(m_2| m_1, \Lambda) \propto 
    \begin{cases}
        0 & m_g \leq m_2 \leq m_g + w_g \, , \\
        m_2^{\beta_q} & \text{otherwise} \, .
    \end{cases}
\end{align}
Here, $m_g$ is the lower boundary of the $m_2$ mass gap and $w_g$ is the width of the gap. We assume a Bayesian prior on $m_g$ that is uniform on the interval $(20 M_\odot, 150 M_\odot)$ while $w_g$ is uniform on the interval $(0 M_\odot, 150 M_\odot)$.
This model implies that the secondary mass is a power-law distribution unless $m_2$ is on the interval $(m_g, m_g + w_g)$ where no mergers are allowed.

We perform the analysis using the \textsc{Broken Power Law + Two Peaks} model for $\pi(m_1)$ from Ref.~\cite{GWTC_4_population}.
In order to investigate the dependence of our results on this model, we then repeat the analysis using the \textsc{Single Power Law + Two Peaks} model from Ref.~\cite{GWTC_4_population}. 
For our initial analysis---before we model the spin transition mass from Ref.~\cite{antonini2024} (below)---we employ the default spin model from Ref.~\cite{GWTC_4_population}, which consists of independent and identical truncated Gaussian distributions for the spin magnitudes of the primary and secondary black holes.
Meanwhile, the cosine of the spin tilt distribution is modelled as a mixture of a uniform distribution and a Gaussian distribution with free mean and width.
The evolution of the merger rate over redshift is modelled as a power-law distribution~\citep{Fishbach:2018edt, gwtc2_pop}.

While \textsc{Broken Power Law + Two Peaks} model is found to be the statistically favoured in Ref.~\cite{GWTC_4_population}, the break mass is not constrained at all when we introduce the flexibility of a gap in $m_2$ distribution. The slopes of the two bands of the broken power law distribution are consistent with each other. We interpret the initial requirement of a second power law in the $m_1$ distribution in current data as a projection of the dearth of high $m_2$ events. Instead, we find that the \textsc{Single Power Law + Two Peaks} model has the highest Bayesian evidence when we allow for a gap in the secondary mass distribution. 
When we employ the \textsc{Broken Power Law + Two Peaks} model, the existence of the $m_2$ mass gap in the model is preferred over the no-gap model by a $\ln {\cal B}$ of ${\sim}3.7$.
Using the \textsc{Single Power Law + Two Peaks} model, the model with a gap is preferred with ${\sim} 84$. 
The plots in this paper use our best mass model, \textsc{Single Power Law + Two Peaks} with a gap, unless otherwise stated. We summarize the model comparison result in Extended Data Table~\ref{tab:model_comparison}.

{
\renewcommand{\figurename}{Extended Data Table}
\setcounter{figure}{0}

\begin{figure}
\centering
\includegraphics[width=1\linewidth]{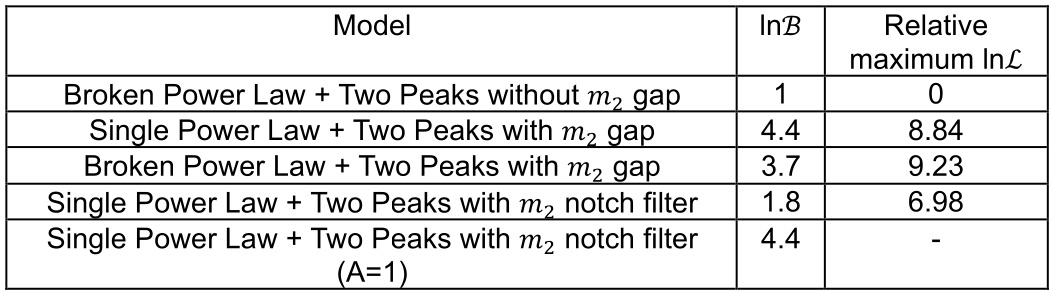}
\caption{The natural logarithm Bayes factor and maximum likelihood comparison of a few variations of the basic mass model with the default model of Ref~\citep{GWTC_4_population} as the baseline. The Bayes factor of Single Power Law + Two Peaks with $m_2$ notch filter model assuming $A = 1$ (i.e., empty gap case) is approximated by the Savage-Dickey density ratio.}
\label{tab:model_comparison}
\end{figure}

}

Extended Data Figure~\ref{fig:corner_m2gap} shows the posterior distribution for the lower edge of the gap $m_g$ and the upper edge of the gap $m_g + w_g$.
The results from GWTC-4, including the exceptionally massive event GW231123 \citep{GW231123}, are shown in blue.
The inclusion of GW231123 is necessary in order to obtain a useful constraint on the upper edge of the gap.
However, as noted before, the large spins of GW231123 may suggest that it is a contamination-event from a 2G+2G merger as opposed to a binary with black holes formed above the pair instability gap.
The constraints on the $m_g$ obtained without GW231123 (orange) are not significantly different from the results obtained with GW231123.
The results obtained from GWTC-3 \citep{gwtc-3} are consistent with results from GWTC-4, albeit with decreased statistical significance.

\setcounter{figure}{0}

\begin{figure}[htp]
    \centering
    \includegraphics[width=1\linewidth]{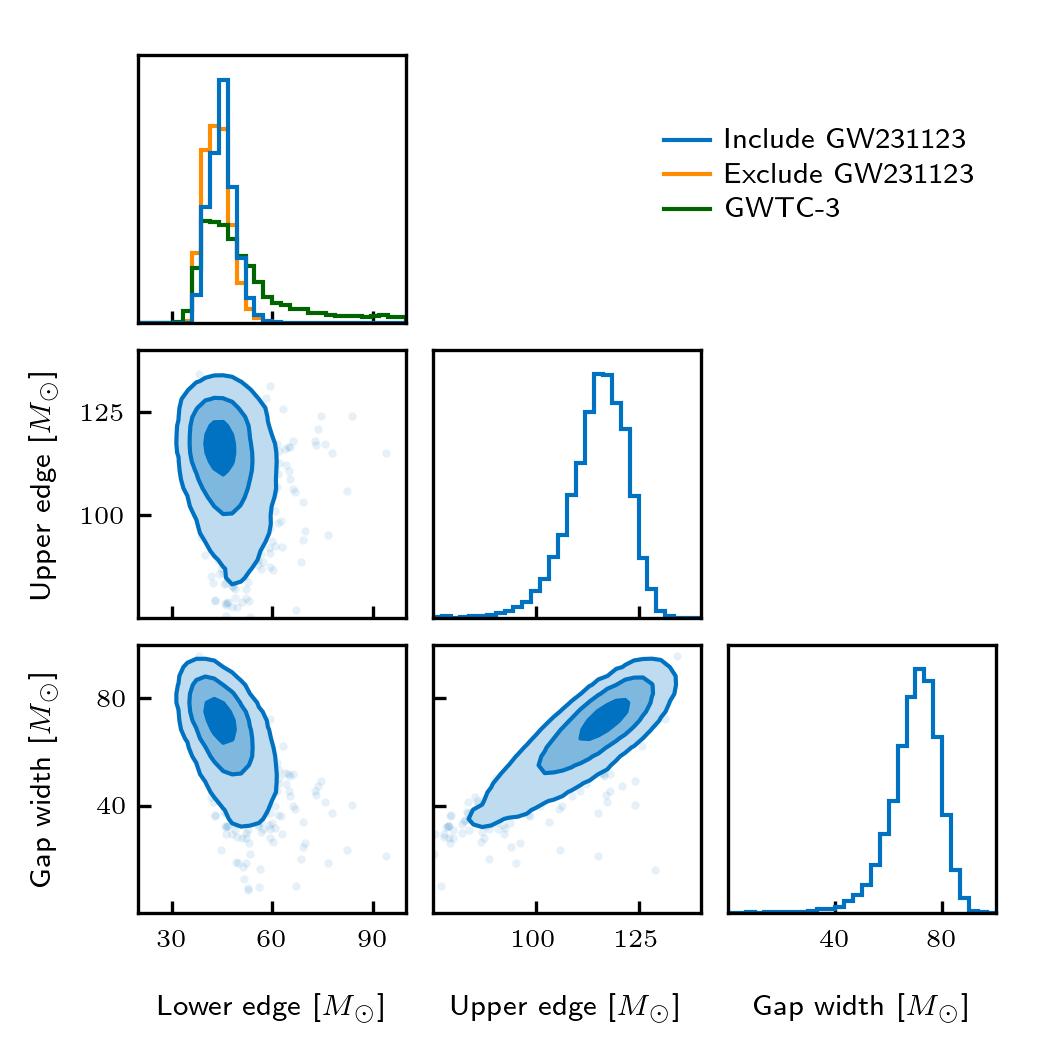}
    \caption{Corner plot of the lower, upper boundary and the width of the secondary mass gap.
    The results from GWTC-4 including the exceptionally massive event GW231123 are shown in blue.
    This event is necessary to obtain a useful constraint on the upper edge of the mass gap.
    Constraints on the lower edge obtained with GWTC-4 excluding GW231123 (orange) and with GWTC-3 (green) are made by inferring the maximum truncation in the $m_2$ distribution.
    The constraints on the lower edge excluding GW231123 are consistent with those obtained with GW231123.
    }
    \label{fig:corner_m2gap}
\end{figure}

\begin{figure}[htp]
    \centering
    \includegraphics[width=1\linewidth]{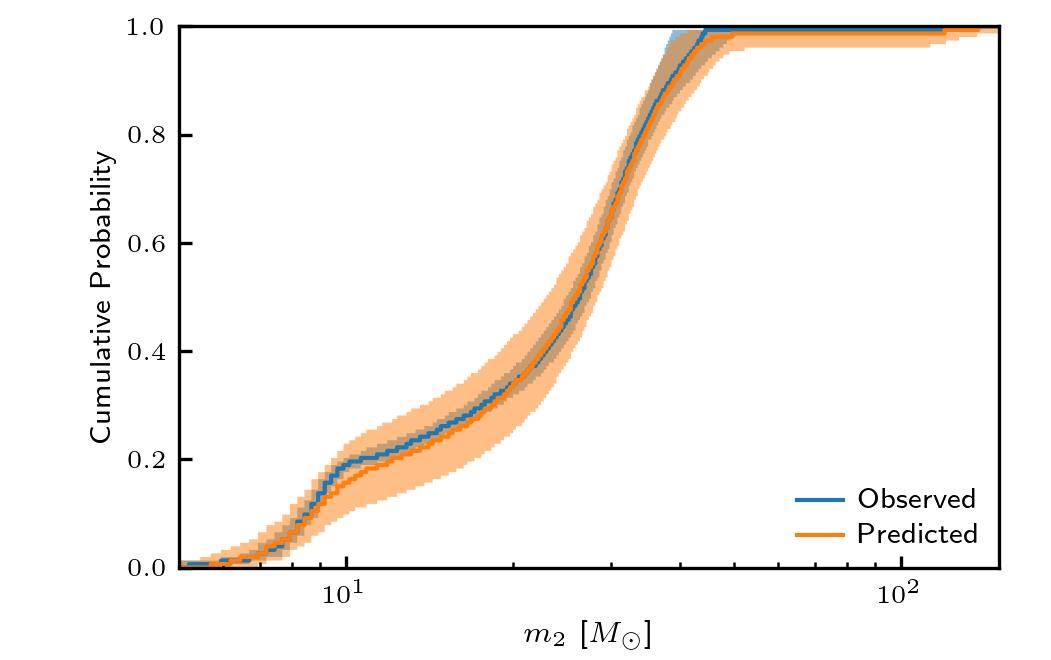}
    \caption{A posterior predictive check. The plots shows the cumulative function of the observed events' secondary-mass distribution (blue) versus the predicted secondary-mass distribution for the \textsc{Single Power Law + Two Peaks} model with a gap in $m_2$ (orange). The solid lines indicate the median and the shaded bands indicate the 90\% credibility range.
    }
    \label{fig:posterior_predictive_check}
\end{figure}

We perform a posterior predictive check~\citep{wmf, Miller:2024sui}.
The distribution predicted by our mass-gap model is consistent with the observed gravitational-wave events, as shown in Extended Data Figure~\ref{fig:posterior_predictive_check}.
The cumulative density of secondary masses $m_2$ reaches a plateau around the lower edge of the gap ${\sim} 45\,M_\odot$ until ${\sim} 116\,M_\odot$ with a tiny bump for both the observed and predicted distributions due to the mergers at the far end of the gap.

\subsection{Gap depth investigation with a notch-filter model}\label{sec:notch_filter}

We replace the model of a fixed empty gap in Eq.~\eqref{eq:gap_model} by a notch filter $n(m_2)$ to investigate the depth of the gap:
\begin{align}\label{eq:notch_model}
    \pi(q | \Lambda) \propto 
        q^{\beta_q} n(m_2).
\end{align}
The expression for the notch filter can be written as \citep{Fishbach:2020ryj} 
\begin{align}\label{eq:notch_expression}
    n(m_2) = 1-\frac{A}{(1+(\frac{m_g}{m_2})^{\eta_{\rm{low}}})(1+(\frac{m_2}{m_g+w_g})^{\eta_\mathrm{high}})} ,
\end{align}
where $m_g$ and $w_g$ are hyper-parameters describing the lower edge and the width of the gap, $\eta_{\rm{low}}$ and $\eta_{\rm{high}}$ set the sharpness of the gap’s edges, and the amplitude of the gap is determined by the parameter $A$. We choose the same priors for $m_g$ and $w_g$ as our standard gap model. We employ uniform priors on $A$ in range of [0,1] as well as $\eta_{\rm{low}}$ and $\eta_{\rm{high}}$ in range of [0,50]. Higher values of  $\eta_{\rm{low}}$ and $\eta_{\rm{high}}$ corresponds to sharp low and high edges respectively.

\begin{figure}[htp]
    \centering
    \includegraphics[width=1\linewidth]{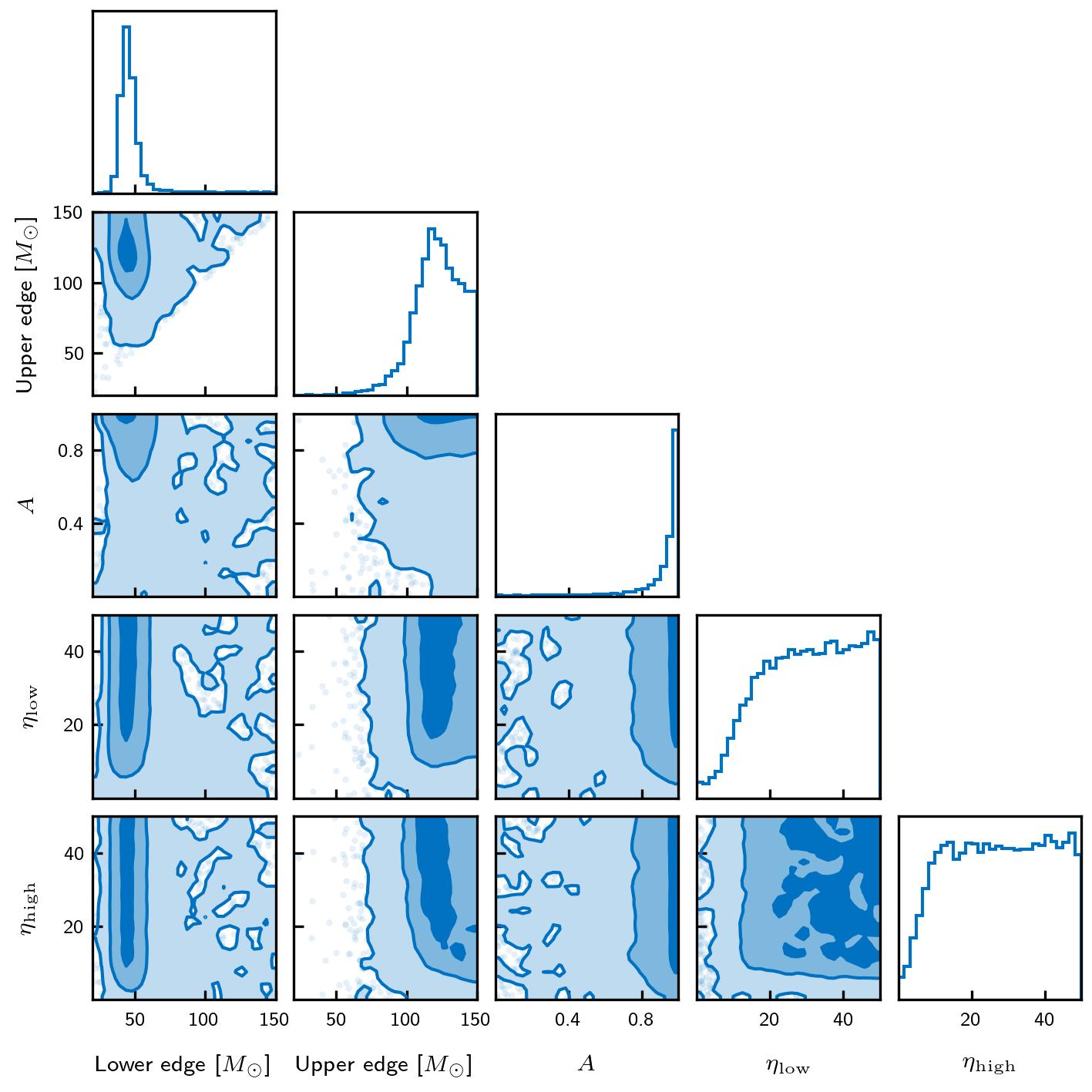}
    \caption{Corner plot of the notch filter parameters. The plots shows the posterior distribution of the parameters that describe the notch filter in Eq.~\eqref{eq:notch_expression}. The contours in the 2D panels indicate the $1\sigma$, $2\sigma$, and $3\sigma$ credible intervals.
    }
    \label{fig:corner_notch}
\end{figure}

We show the posteriors of those parameters in Extended Data Figure~\ref{fig:corner_notch}. The posterior of A is sharply peaked at 1, which represents the preference of an empty gap with the lower edge of the gap consistent with our default model result. The inference of the upper edge in this notch filter model is much more uncertain given a more flexible description of the gap, which is expected since it is mostly driven by the detection of a single event. Non-sharp edges are disfavoured, in particularly for the lower end, and increasing sharpnesses are not distinguishable (also from a hard cutoff) given current dataset.

With the flexibility of a finite depth of the gap in this model, we can also place an upper limit on the rate of binary black holes that have both components inside the gap by the non-detection of such events. Assuming that any component inside the gap is a 2G black hole, we place constraints on the upper limit of the merger rates of 2G+2G black holes in the gap to be $< 7.9\times 10^{-2}\,\rm{Gpc}^{-3}\,\rm{yr}^{-1}$ at 90\% credibility.

\subsection{Comparison to the maximum population likelihood distribution}\label{sec:pi_stroke}

The $\pistroke$ formalism introduced in Ref.~\cite{PhysRevResearch.5.023013} offers a data-driven approach to visualizing the population properties of merging binaries.
Rather than fitting a model, the $\pistroke$ formalism identifies the unique distribution that maximizes the population likelihood over all possible population models. The resulting distribution $\pistroke(\theta)$ is always given by a weighted sum of delta functions, with the locations and weights of the ``$\pistroke$ samples'' are determined via constrained numerical optimization. 
The samples show which population features are supported by the data, as opposed to features that might be inferred due to model-dependence. 
In the limit where the number of observations goes to infinity, the $\pistroke$ samples reproduce the true distribution \cite{PhysRevResearch.5.023013}.

\begin{figure}[htbp]
    \centering
    \includegraphics[width=1\textwidth]{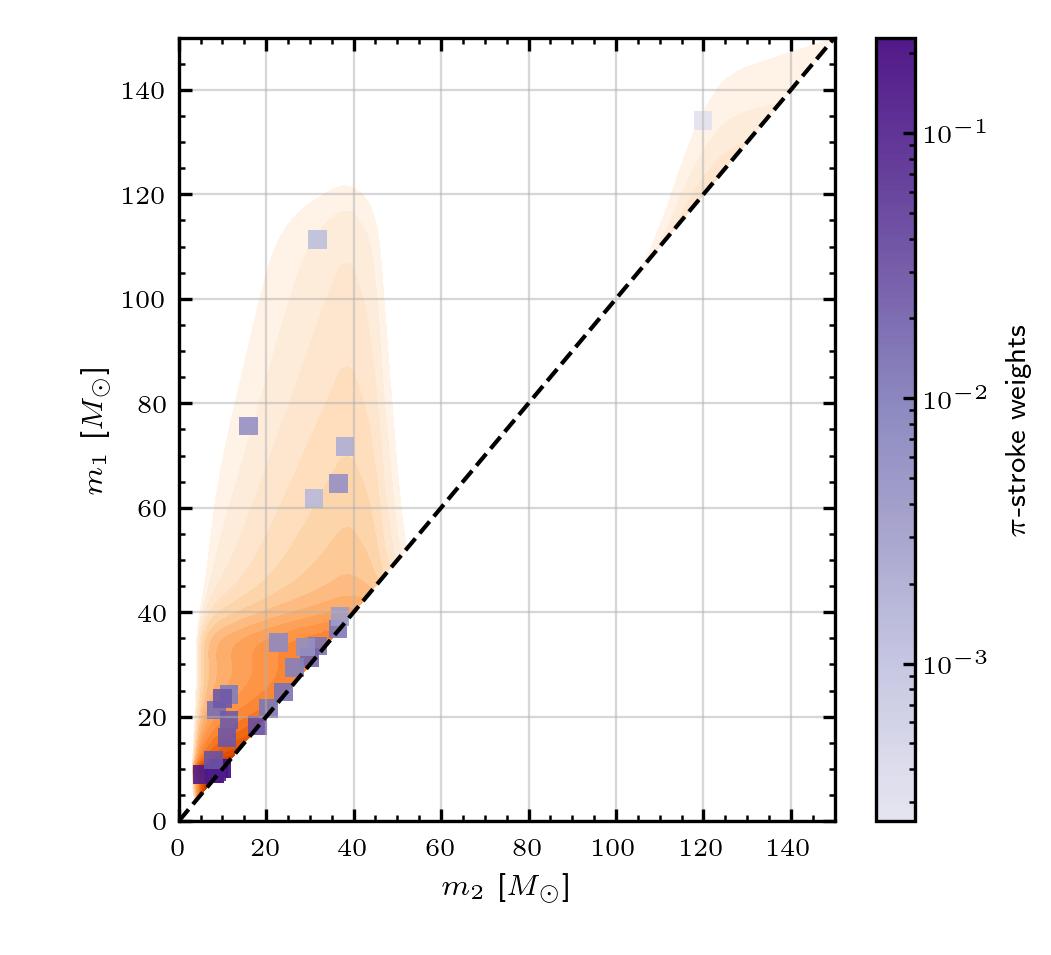}
    \caption{Predicted population distribution by maximum population likelihood method, the $\pi$-stroke (\st{$\pi$}) formalism. The distribution is expressed as a weighted sum of delta functions. Scatter points show where the delta functions are in terms of ($m_1$, $m_2$). The colorbar represents the weights of delta functions. We also show the mean prediction of the joint distribution $\pi(m_1, m_2)$ by our standard parametric model analysis in the background for reference. Both analyses show consistent support for $m_2$ gap.
}
    \label{fig:m1m2_pi_stroke}
\end{figure}

Extended Data Figure~\ref{fig:m1m2_pi_stroke} shows \st{$\pi$} in $(m_1, m_2)$ for the data in GWTC-4. The background contour shows the probability density inferred by the standard parametric model, \textsc{Single Power Law + Two Peaks} with $m_2$ gap. 
The $\pistroke$ samples are consistent with our fit; there is a notable absence of $\pistroke$ samples with secondary mass between approximately $45\,M_\odot$ and $116\, M_\odot$. 
The $\pistroke$ gap is not imposed by any modelling assumptions, but rather emerges organically from the data. 
The agreement between the model-free $\pistroke$ distribution and the parametric model analysis gives us additional confidence in the existence of a pair-instability mass gap in the distribution of secondary component masses.

\subsection{Mass model misspecification test}
As a further test of model misspecification, we also run the analysis using the pairing mass model from Ref.~\cite{Farah:2023swu}.
In this model, the joint distribution for $m_1$ and $m_2$ is written as
\begin{align}
    \pi(m_1, m_2) \propto \pi(m_1 | \Lambda_1) \, 
    \pi(m_2 | \Lambda_2) \, 
    f(q | \beta_q) .
\end{align}
The hyper-parameters $\Lambda_1$ control the shape of a nominal distribution for primary mass while the hyper-parameters $\Lambda_2$ control the shape of a nominal distribution of secondary mass (we use the word ``nominal'' here because the marginal distribution $\int dm_2 \, \pi(m_1, m_2)$ is not equal to $\pi(m_1|\Lambda_2)$).
The pairing function $f(q | \beta_q)$ is a power law in mass ratio $q=m_2/m_1$.
With more parameters than our \textsc{Single Power Law + Two Peaks} model, this pairing mass model provides significantly more flexibility to reconstruct distributions in the $m_1$--$m_2$ plane.

\begin{figure}[htp]
    \centering
    \includegraphics[width=1\linewidth]{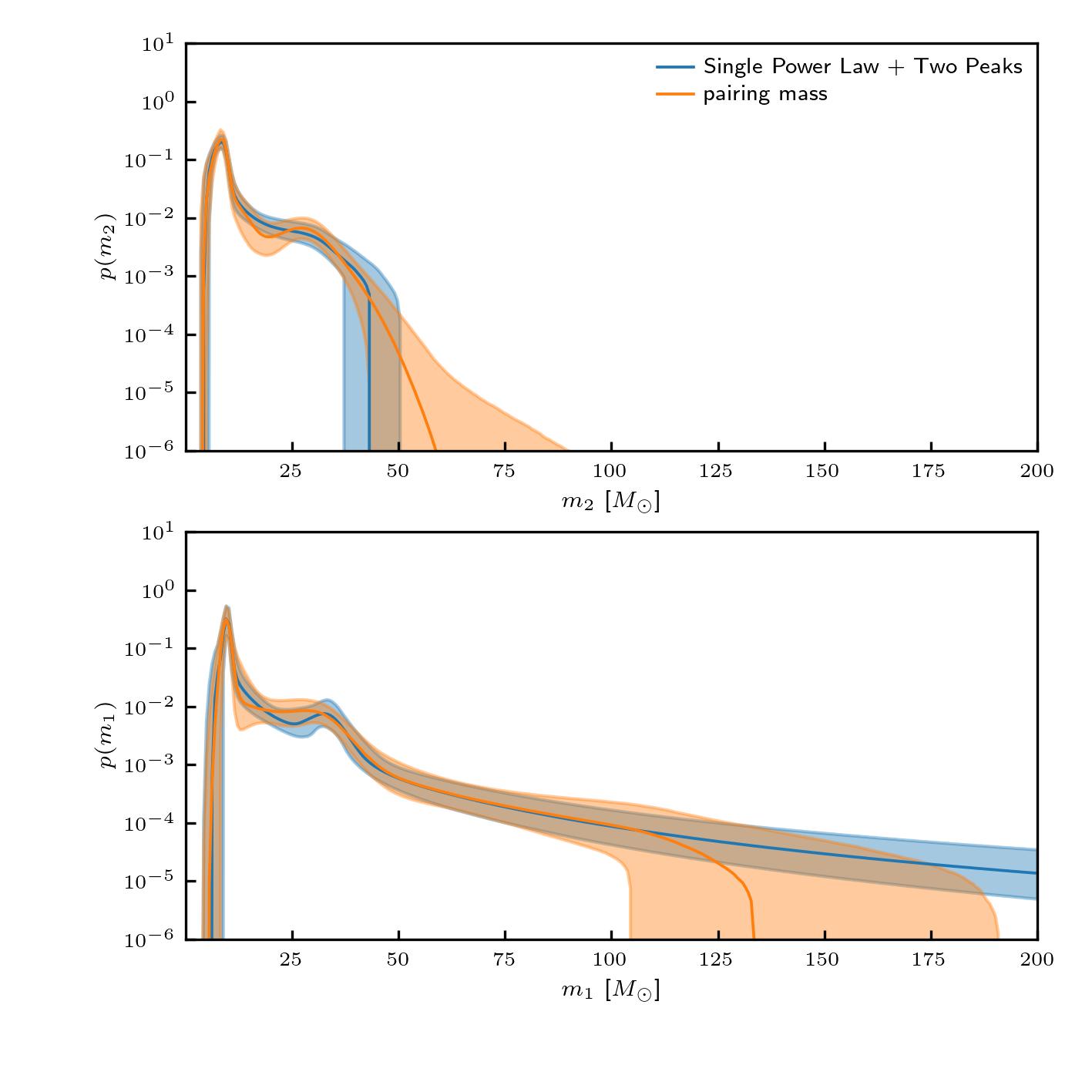}
    \caption{Reconstructed distributions of primary masses $m_1$ (top) and secondary masses $m_2$ (bottom). The blue fit is obtained with the \textsc{Single Power Law + Two Peaks} model while the orange fit is obtained with the pairing mass model from Ref.~\citep{Farah:2023swu}. The shaded bands indicate the 90\% credibility ranges.
    }
    \label{fig:pairing_function_reconstruction}
\end{figure}

\begin{figure}[htp]
    \centering
    \includegraphics[width=1\linewidth]{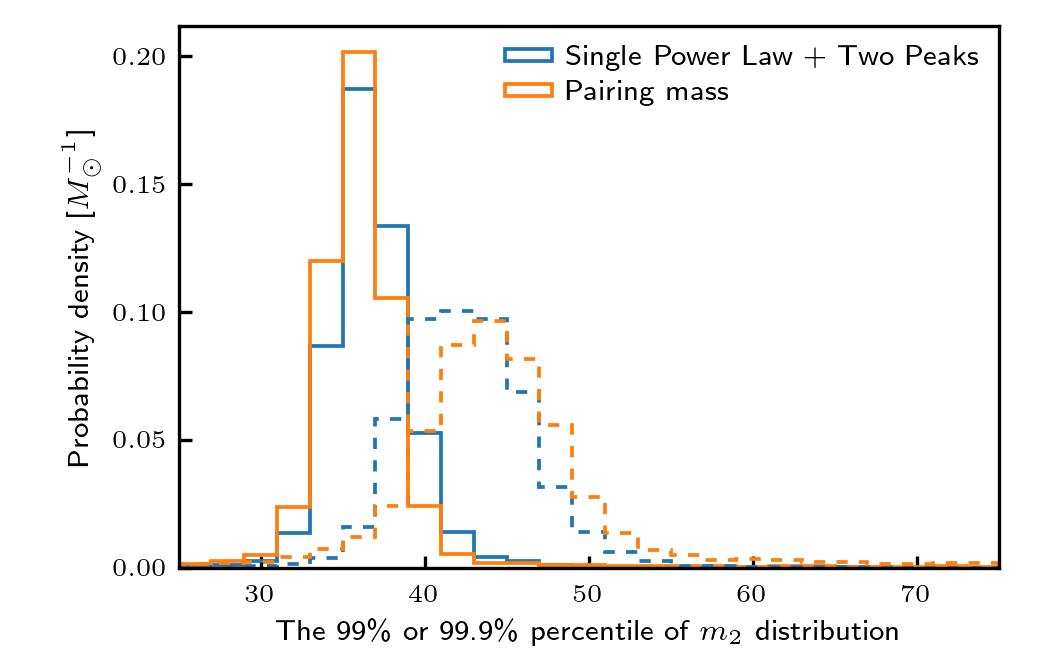}
    \caption{
    Comparing quantiles for different models: 99\% of mergers are below $m_2^{99\%}$ (solid) and 99.9\% are below $m_2^{99.9\%}$ (dashed-dotted).
    Blue shows the results for the \textsc{Single Power Law + Two Peak} model while orange shows the results for the more flexible pairing mass model\citep{Farah:2023swu}.
    }
    \label{fig:m2_percentile_comparison}
\end{figure}

In our version of the pairing mass model, we employ separate \textsc{Broken Power Law + Two Peaks} models for $\pi(m_1|\Lambda_1)$ and $\pi(m_2|\Lambda_2)$.
For the sake of simplicity of a test focusing on the lower edge of the gap, we exclude GW231123.
We additionally allow for the presence of different maximum truncation for $m_1$ and $m_2$.
The reconstructed distributions are shown in Extended Data Figure~\ref{fig:pairing_function_reconstruction}. 
Interestingly, the pairing mass model does not require hard cutoff in the $m_2$ distribution around $45M_\odot$.
This is because it can produce a dearth of events with $m_2 \gtrsim 45 M_\odot$ with a steeply falling power law.
This is evidenced by comparing upper quantiles for the pairing mass and \textsc{Single Power Law + Two Peaks} mass models.
In Extended Data Figure~\ref{fig:m2_percentile_comparison}, we plot the 99th percentile $m_{2}^{99\%}$ of secondary masses in the population.
The two distributions from \textsc{Single Power Law + Two Peaks} model and the pairing mass model agree.
Repeating this exercise for the $99.9\%$ quantile (dashed), we see the two distributions again agree.
We conclude that the pairing mass model fit is \textit{functionally equivalent} to one with a hard cutoff in $m_2$, even though this can be achieved without an explicit cutoff.
The difference in $m_1>175 M_\odot$ region between the models is caused by a fixed maximum $m_1=300M_\odot$ truncation in our \textsc{Single Power Law + Two Peaks} model following the setup of Ref.~\cite{GWTC_4_population}.
However, the maximum $m_1$ truncation is not constrained at all when it is analysed as a free parameter. The choice of a fixed maximum $m_1$ truncation makes no difference in the inference of other major structures of the mass distribution.

\subsection{Spin transition study}\label{sec:spin_model}

\begin{figure}[htp]
    \centering
    \includegraphics[width=1\linewidth]{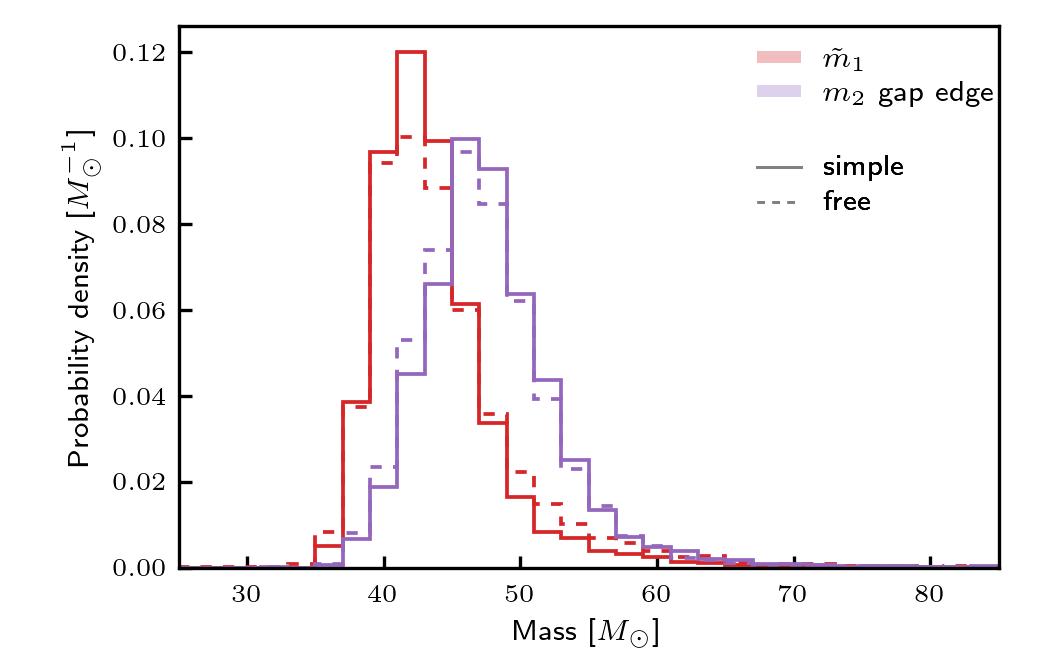}
    \caption{
    Comparing the lower edge of the mass gap and the spin transition mass results using different spin models.
    We show the results in solid lines assuming the ``simple'' spin transition model from Eq.~\eqref{model:spin_simple} while the results in dashed-dotted assume the more complicated spin transition model from Eq.~\eqref{model:spin_complex}. In purple is the posterior for the lower edge of the $m_2$ mass gap. Red shows the posterior for the spin transition mass $\tilde{m}_1$.
    The results are consistent between different spin transition models.
    }
    \label{fig:spin_model_comparison}
\end{figure}

Next, we replace the default spin model with the model from Ref.~\cite{antonini2024}, which showed evidence for a change in the spin distribution for primary masses above $\tilde{m}_1$.
In our study allowing an $m_2$ gap, we consider two model variations of the effective inspiral spin $\chi_{\rm{eff}}$, \begin{align} \chi_\text{eff} \equiv \frac{\cos(\text{tilt}_1)\chi_1 + q\cos(\text{tilt}_2)\chi_2}{1+q}. \end{align}
In the first variation, the effective inspiral spin parameter $\chi_{\rm{eff}}$ is normally distributed when $m_1 < \tilde{m}_1$ and it is uniformly distributed when $m_1 > \tilde{m}_1$ \citep{antonini2024}:
\begin{equation}\label{model:spin_simple}
\begin{aligned}
\pi(\chi_\mathrm{eff}|m_1,\Lambda) & =
\begin{cases}
\mathcal{N}(\chi_\mathrm{eff}|\mu,\sigma) & (m_1<\tilde{m}_1) \\
\mathcal{U}(\chi_\mathrm{eff}|w=0.47) & (m_1 \geq \tilde{m}_1).
\end{cases}
\end{aligned}
\end{equation}
The second variation employs a mixture model for binaries with $m_1 > \tilde{m}_1$\citep{antonini2024}:
\begin{equation}\label{model:spin_complex}
\begin{aligned}
\pi(\chi_\mathrm{eff}|m_1,\Lambda) & =
\begin{cases}
\mathcal{N}(\chi_\mathrm{eff}|\mu,\sigma) & (m_1<\tilde{m}_1) \\
\zeta\,\mathcal{U}(\chi_\mathrm{eff}|w) + (1-\zeta) \mathcal{N}_{\rm u}(\chi_\mathrm{eff}|\mu_{\rm u}, \sigma_{\rm u}) & (m_1 \geq \tilde{m}_1).
\end{cases}
\end{aligned}
\end{equation}
We show in Extended Data Figure~\ref{fig:spin_model_comparison} that both models produce similar results.
However, the complexity of Eq.~\eqref{model:spin_complex} means this model is disfavoured due to an Occam penalty.
We therefore report our results using the spin model Eq.~\eqref{model:spin_simple} throughout the paper, unless otherwise specified. As Ref.~\citep{antonini2024} showed, the transition-spin model (without an $m_2$ gap) is strongly favoured over a model in which the entire population is represented by a single Gaussian in $\chi_{\rm{eff}}$ by ${\ln \cal{B}} > 9$. Our fiducial model, in which the spin transition mass is equal to the lower edge of the $m_2$ gap, is further favoured over Ref.~\citep{antonini2024}'s model ($ {\ln \cal{B}} {\sim} 3.3$), as well as a model in which the spin transition mass can be different from the lower edge of the $m_2$ gap (${\ln \cal{B}} {\sim} 1.0$).

Ref.~\citep{antonini2025} further investigates this spin transition feature using a non-parametric model for the spin distribution of the high mass sub-population. It is shown that with a weaker assumption on the spin model, the measurement of the spin transition mass $\tilde{m}_1$ is still consistent with the results using the parametric models introduced above.

\subsection{Symmetry of the effective spin distribution for the high mass mergers}\label{sec:spin_symmetry}
We explore whether the spin distribution for the high mass mergers is isotropic or not by a more flexible model following ~\citep{antonini2025, antonini2025_PISN}. 
We extend model in Eq.~\ref{model:spin_simple} to allow the minimum and maximum truncation of the uniform distribution which describes the spin distribution for the high mass mergers to be free parameters. 
\begin{equation}\label{model:spin_flexible}
\begin{aligned}
\pi(\chi_\mathrm{eff}|m_1,\Lambda) & =
\begin{cases}
\mathcal{N}(\chi_\mathrm{eff}|\mu,\sigma) & (m_1<\tilde{m}_1) \\
\mathcal{U}(\chi_\mathrm{eff}|\chi_\text{min}^\text{high-mass}, \chi_\text{max}^\text{high-mass}) & (m_1 \geq \tilde{m}_1).
\end{cases}
\end{aligned}
\end{equation}
We set the prior for $\chi_\text{min}^\text{high-mass}$ that is uniform on the interval $(-1, 1)$ and $\chi_\text{max}^\text{high-mass}$ that is uniform on the interval $(-1, 1)$. 

\begin{figure}[htp]
    \centering
    \includegraphics[width=1.0\linewidth]{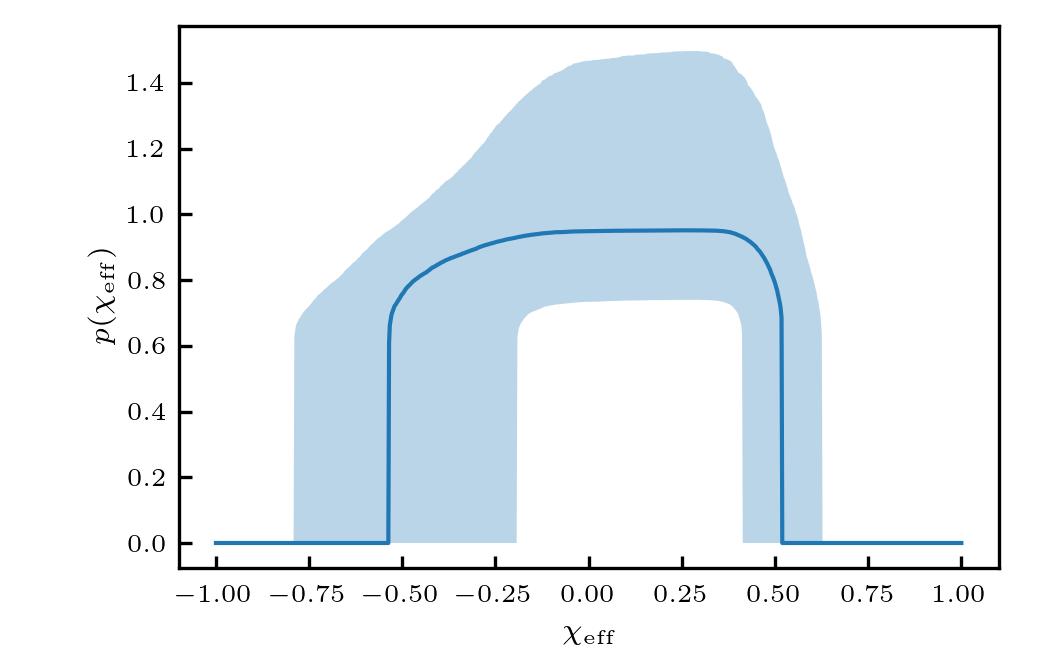}
    \caption{The effective spin distribution for high mass binary black holes with $m_1>\tilde{m}_1$ using the flexible spin model in Eq.~\ref{model:spin_flexible}. The shaded area represents the 90\% credible interval and the solid line is the median.}
    \label{fig:high_mass_spin_flexible}
\end{figure}

We show the result of the spin distribution for high mass mergers in Extended Data Figure~\ref{fig:high_mass_spin_flexible}. While the lower boundary of the spin distribution is relatively less constrained than the upper boundary, the overall distribution is consistent with a spin distribution symmetric around zero. We found the mean $\langle\chi_\text{eff}\rangle=((\chi_\text{max}^\text{high-mass}+\chi_\text{min}^\text{high-mass})/2=-0.00^{+0.12}_{-0.10}$ and the half-width $\rm{w}^{\chi}_{\mathrm{high-mass}}=((\chi_\text{max}^\text{high-mass}-\chi_\text{min}^\text{high-mass})/2=0.53^{+0.12}_{-0.15}$. A positive minimum truncation, $\chi_\text{min}^\text{high-mass}>0$, is ruled out at 99.3\% credibility.

\subsection{Constraints on the $^{12}\rm{C}(\alpha,\gamma)^{16}\rm{O}$ rate with different models}\label{sec:S-factor}

Based on simulations, Ref.~\cite{Farmer:2020xne} fit the lower edge of the mass gap as a function of the temperature-dependent uncertainty in the $^{12}\rm{C}(\alpha,\gamma)^{16}\rm{O}$ rate and reported constraints on the $^{12}\rm{C}(\alpha,\gamma)^{16}\rm{O}$ using the first ten gravitational-wave detections.
However, in order to determine the lower edge of the mass gap with so few events, the analysis had to employ a model with rather strong assumptions about the shape of the black hole mass distribution.
Following the observation of additional binary black hole signals, it became apparent that this early model was misspecified, and so the earliest inferences on the lower edge of the mass gap were unreliable~\citep{gwtc2_pop,gwtc3_pop}. 
A more recent analysis \citep{Golomb:2023vxm} explored the possibility of the bump around $35\,M_\odot$ in binary black hole mass distribution to be a signature from the pulsational pair-instability process and found the inferred astrophysical $S$-factor in tension with Ref.~\cite{deBoer_2017}, illustrating the difficulty of placing constraints on $^{12}\rm{C}(\alpha,\gamma)^{16}\rm{O}$ reaction rates using previous gravitational-wave catalogues.

\begin{figure}[htp]
    \centering
    \includegraphics[width=1.0\linewidth]{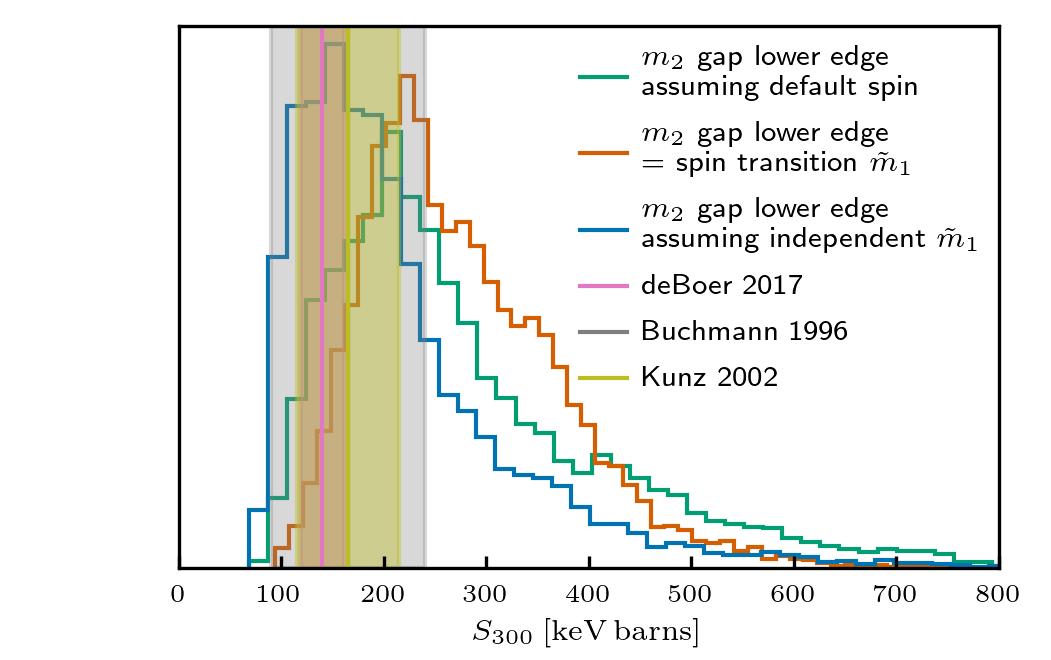}
    \caption{
    Constraints on $^{12}\rm{C}(\alpha,\gamma)^{16}\rm{O}$ rate with various population model assumptions.
    Green is the posterior for $S_{300}$ using the lower edge of the mass gap inferred assuming default spin model in Ref.~\citep{GWTC_4_population}. Blue uses the measurement of the lower edge of the mass gap assuming the spin transition model in Ref.~\citep{antonini2024} while the spin transition mass $\tilde{m}_1$ is independent and non-identical of the $m_2$ gap. 
    Orange is the posterior using the measurement imposing lower edge of the $m_2$ gap equal to the spin transition mass $\tilde{m}_1$.
    Results in pink, grey and and yellow-green show the theoretical nuclear physics predictions from Refs.~\cite{deBoer_2017, Buchmann_1996, Kunz_2002}.}
    \label{fig:all_nuclear physics}
\end{figure}

Here, we show the constraints on the $^{12}\rm{C}(\alpha,\gamma)^{16}\rm{O}$ rate with other model assumptions in addition to the one in the main body of the paper. As mentioned, we used the relationship between the lower edge of the mass gap and the $^{12}\rm{C}(\alpha,\gamma)^{16}\rm{O}$ rate fit from Ref.~\cite{Farmer:2020xne}.
As shown in Extended Data Figure~\ref{fig:all_nuclear physics}, we constrain the astrophysical $S$-factor of $^{12}\rm{C}(\alpha,\gamma)^{16}\rm{O}$ at $\unit[300]{keV}$ to $\unit[243_{-119}^{+331}]{keV \, barns}$ (90\% credibility) assuming the default spin model and to $\unit[260_{-108}^{+190}]{keV \, barns}$ (90\% credibility) by incorporating spin transition information. In blue we also show the constraint to be $\unit[186_{-88}^{+282}]{keV \, barns}$ using the mass gap measurement which assumes that the spin transition mass is independent of the $m_2$ gap. The measurements of the mass gap shift slightly given different assumptions on spin models due to the degeneracy, which leads to different but consistent posteriors of the astrophysical $S$-factor.

Since the mapping from the lower edge of the gap to the corresponding $S$-factor is non-linear, small shifts towards a lower inferred edge lead to a long tail towards higher inferred $S$-factors. We see that using a different model that shifts the lower edge by $\sim 2\,M_\odot$ results in a $S$-factor estimate that aligns well with the result of Ref.~\citep{deBoer_2017} (see the blue curve of Extended Data Figure~\ref{fig:all_nuclear physics}). 
Moreover, the simulations mapping from the gap to the corresponding $^{12}\rm{C}(\alpha,\gamma)^{16}\rm{O}$ rate in Ref.~\cite{Farmer:2020xne} simulated the lowest edge $\sim 40 M\,_\odot$ while our constraint has an error bar extending to $< 40 M\,_\odot$. Thus, the long tail to high $S$-factors in our posterior may be unrealistic.

\subsection{Implications for GW190521}\label{sec:GW190521}

\begin{figure}[htp]
    \centering
    \includegraphics[width=1\linewidth]{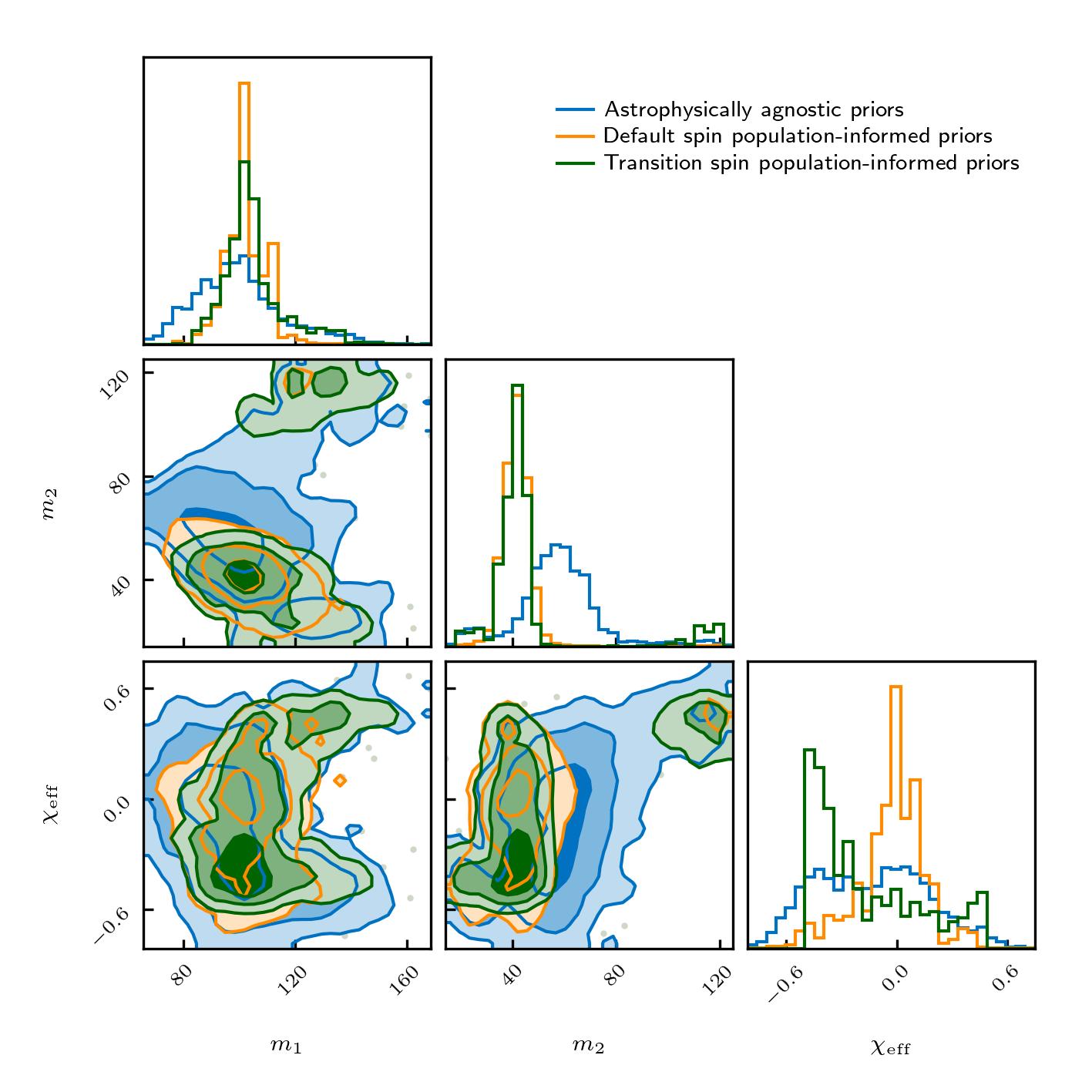}
    \caption{The inferred population-informed estimates of the primary mass $m_1$, secondary mass $m_2$, and effective spin $\chi_\mathrm{eff}$ for GW190521.
    The contours represent $1\sigma$, $2\sigma$, $3\sigma$ credible regions respectively.
    }
    \label{fig:corner_GW190521}
\end{figure}

The presence of an apparent gap due to pair-instability supernovae has interesting implications for previously identified gravitational-wave events with exceptional properties. GW190521 was highlighted as a potential candidate of hierarchical mergers in the mass gap \citep{GW190521, GW190521_implications}. 
While this interpretation is consistent with our results, we also find support for the hypothesis---first put forward in Ref.~\cite{Fishbach_GW190521}---that GW190521 is a straddling binary with component black holes on either side of the gap. 
We show a corner plot for the mass and spin properties of GW190521 in Extended Data Figure~\ref{fig:corner_GW190521}. 
Within the context of our population model~\citep{2020PhRvD.102h3026G, 2020ApJ...891L..31F}, the secondary black hole in GW190521 is most likely below the pair instability gap: $m_2=40.7^{+8.9}_{-7.0}\,M_\odot$ at 90\% credibility, though there is still marginal support for $m_2$ at the other end of the gap ${\sim}120\,M_\odot$. There is posterior support $\sim 4\%$ for the hypothesis that both components of GW190521 are on the far side of the gap. Excluding GW231123, the overall population has a maximum $m_2$ $\sim 43\,M_\odot$ as shown in orange in Extended Data Figure~\ref{fig:corner_m2gap}. Thus the event GW190521 in this case is fully consistent with its secondary component below the gap.

For future work, we envision a global fit using a model that includes subpopulations of 1G+1G, 2G+1G and 2G+2G mergers; see, e.g., Refs.~\citep{Doctor_2020,hierarchical,Kimball:2020qyd,Mould:2022ccw, Wang:2022gnx}.
By explicitly modeling the 2G+1G subpopulation, it would be possible to account for contamination from hierarchical mergers in the $m_1$ mass gap.
As the number of detections increases, it will be possible to gain new insights into the pair-instability gap and the prevalence of hierarchical mergers in merging binaries.

{ \ }

{\bf Data Availability.} Results of the analyses in this work can be found on Zenodo\citep{mass_gap_zenodo}. The posterior samples of GWTC-4 events used in the hierarchical Bayesian inference of this work are available on Zenodo: 16053484 as part of LIGO-Virgo-Kagra's GWTC-4 data release \citep{GWTC_4_PE_release}. For events in GWTC-2.1 and GWTC-3, we use samples from Zenodo: 6513631 and Zenodo: 554666 \citep{GWTC_2.1_PE_release,GWTC_3_PE_release}.
We use the cumulative search sensitivity file on Zenodo: 16740128 \citep{GWTC_4_sensitivity_release}.

{ \ } 

{\bf Code Availability.} 

The code for this study is publicly available at \url{https://github.com/HuiTong5/PISN_mass_gap_GWTC-4} as an implementation of the population models developed in this study using \texttt{GWPopulation} \cite{gwpopulation}.

{ \ }

{\bf Acknowledgements}
We thank Zoheyr Doctor for inspiring this study. We thank Simon Stevenson, Christopher Berry, Thomas Dent, Anjali Yelikar, Parthapratim Mahapatra, Isobel Romero-Shaw, Fabio Antonini, Elenna Capote and Vaibhav Tiwari for comments.
H.T. and E.T. are supported by the Australian Research Council CE230100016, LE210100002, DP230103088.
M.F. acknowledges support from the Natural Sciences and Engineering Research Council of Canada (NSERC) under grant RGPIN-2023-05511, the University of Toronto Connaught Fund, the Alfred P. Sloan Foundation, and the Ontario Early Researcher Award. 
M.M. is supported by a Research Fellowship from the Royal Commission for the Exhibition of 1851 and by the LIGO Laboratory through the National Science Foundation awards PHY-1764464 and PHY-2309200.

This material is based upon work supported by NSF's LIGO Laboratory which is a major facility fully funded by the National Science Foundation.
The authors are grateful for for computational resources provided by the LIGO Laboratory computing cluster at California Institute of Technology supported by National Science Foundation Grants PHY-0757058 and PHY-0823459. LIGO was constructed by the California Institute of Technology and Massachusetts Institute of Technology with funding from the National Science Foundation and operates under cooperative agreement PHY-1764464. This paper carries LIGO Document Number LIGO-P2500453.

{ \ }

\noindent {\bf Author contributions} 

\noindent 
The analyses in this work were carried out by Tong. 
Fishbach and Thrane served as Tong's supervisors, providing guidance and assisting with the writing of the manuscript.
Additional support was provided by Mould, Callister, Farah, and Banagiri, who provided suggestions for analyses and helped interpret the results. Guttman performed the maximum population likelihood distribution analysis. Mould contributed to the initial code development. Galaudage helped with plotting. Beltran-Martinez, Farr, Galaudage, Godrey, Heinzel, Kalomenopoulos, Miller and Vijaykumar contributed to preliminary population inference that motivated this study.
All authors contributed to editing the manuscript.

{ \ }

\noindent {\bf Corresponding author} 

\noindent 
Correspondence to Hui Tong.

{ \ }

\noindent {\bf Competing interests}

\noindent The authors declare that they have no competing interests.


\begin{thebibliography}{49}
\ifx \bisbn   \undefined \def \bisbn  #1{ISBN #1}\fi
\ifx \binits  \undefined \def \binits#1{#1}\fi
\ifx \bauthor  \undefined \def \bauthor#1{#1}\fi
\ifx \batitle  \undefined \def \batitle#1{#1}\fi
\ifx \bjtitle  \undefined \def \bjtitle#1{#1}\fi
\ifx \bvolume  \undefined \def \bvolume#1{\textbf{#1}}\fi
\ifx \byear  \undefined \def \byear#1{#1}\fi
\ifx \bissue  \undefined \def \bissue#1{#1}\fi
\ifx \bfpage  \undefined \def \bfpage#1{#1}\fi
\ifx \blpage  \undefined \def \blpage #1{#1}\fi
\ifx \burl  \undefined \def \burl#1{\textsf{#1}}\fi
\ifx \doiurl  \undefined \def \doiurl#1{\url{https://doi.org/#1}}\fi
\ifx \betal  \undefined \def \betal{\textit{et al.}}\fi
\ifx \binstitute  \undefined \def \binstitute#1{#1}\fi
\ifx \binstitutionaled  \undefined \def \binstitutionaled#1{#1}\fi
\ifx \bctitle  \undefined \def \bctitle#1{#1}\fi
\ifx \beditor  \undefined \def \beditor#1{#1}\fi
\ifx \bpublisher  \undefined \def \bpublisher#1{#1}\fi
\ifx \bbtitle  \undefined \def \bbtitle#1{#1}\fi
\ifx \bedition  \undefined \def \bedition#1{#1}\fi
\ifx \bseriesno  \undefined \def \bseriesno#1{#1}\fi
\ifx \blocation  \undefined \def \blocation#1{#1}\fi
\ifx \bsertitle  \undefined \def \bsertitle#1{#1}\fi
\ifx \bsnm \undefined \def \bsnm#1{#1}\fi
\ifx \bsuffix \undefined \def \bsuffix#1{#1}\fi
\ifx \bparticle \undefined \def \bparticle#1{#1}\fi
\ifx \barticle \undefined \def \barticle#1{#1}\fi
\bibcommenthead
\ifx \bconfdate \undefined \def \bconfdate #1{#1}\fi
\ifx \botherref \undefined \def \botherref #1{#1}\fi
\ifx \url \undefined \def \url#1{\textsf{#1}}\fi
\ifx \bchapter \undefined \def \bchapter#1{#1}\fi
\ifx \bbook \undefined \def \bbook#1{#1}\fi
\ifx \bcomment \undefined \def \bcomment#1{#1}\fi
\ifx \oauthor \undefined \def \oauthor#1{#1}\fi
\ifx \citeauthoryear \undefined \def \citeauthoryear#1{#1}\fi
\ifx \endbibitem  \undefined \def \endbibitem {}\fi
\ifx \bconflocation  \undefined \def \bconflocation#1{#1}\fi
\ifx \arxivurl  \undefined \def \arxivurl#1{\textsf{#1}}\fi
\csname PreBibitemsHook\endcsname


\bibitem[\protect\citeauthoryear{Fowler and Hoyle}{1964}]{Fowler:1964zz}
\begin{barticle}
\bauthor{\bsnm{Fowler}, \binits{W.A.}},
\bauthor{\bsnm{Hoyle}, \binits{F.}}:
\batitle{{Neutrino Processes and Pair Formation in Massive Stars and Supernovae}}.
\bjtitle{Astrophys. J. Suppl.}
\bvolume{9},
\bfpage{201}--\blpage{319}
(\byear{1964})
\end{barticle}
\endbibitem

\bibitem[\protect\citeauthoryear{{Rakavy} and {Shaviv}}{1967}]{1967ApJ...148..803R}
\begin{barticle}
\bauthor{\bsnm{{Rakavy}}, \binits{G.}},
\bauthor{\bsnm{{Shaviv}}, \binits{G.}}:
\batitle{{Instabilities in Highly Evolved Stellar Models}}.
\bjtitle{Astrophys. J.}
\bvolume{148},
\bfpage{803}
(\byear{1967})
\end{barticle}
\endbibitem

\bibitem[\protect\citeauthoryear{Barkat et~al.}{1967}]{1967_PhysRevLett.18.379}
\begin{barticle}
\bauthor{\bsnm{Barkat}, \binits{Z.}},
\bauthor{\bsnm{Rakavy}, \binits{G.}},
\bauthor{\bsnm{Sack}, \binits{N.}}:
\batitle{Dynamics of supernova explosion resulting from pair formation}.
\bjtitle{Phys. Rev. Lett.}
\bvolume{18},
\bfpage{379}--\blpage{381}
(\byear{1967})
\end{barticle}
\endbibitem

\bibitem[\protect\citeauthoryear{{Fraley}}{1968}]{1968Ap&SS...2...96F}
\begin{barticle}
\bauthor{\bsnm{{Fraley}}, \binits{G.S.}}:
\batitle{{Supernovae Explosions Induced by Pair-Production Instability}}.
\bjtitle{Ap\&SS}
\bvolume{2}(\bissue{1}),
\bfpage{96}--\blpage{114}
(\byear{1968})
\end{barticle}
\endbibitem

\bibitem[\protect\citeauthoryear{Heger and Woosley}{2002}]{Heger:2001cd}
\begin{barticle}
\bauthor{\bsnm{Heger}, \binits{A.}},
\bauthor{\bsnm{Woosley}, \binits{S.E.}}:
\batitle{{The nucleosynthetic signature of population III}}.
\bjtitle{Astrophys. J.}
\bvolume{567},
\bfpage{532}--\blpage{543}
(\byear{2002})
\end{barticle}
\endbibitem

\bibitem[\protect\citeauthoryear{Woosley et~al.}{2007}]{Woosley:2007qp}
\begin{barticle}
\bauthor{\bsnm{Woosley}, \binits{S.E.}},
\bauthor{\bsnm{Blinnikov}, \binits{S.}},
\bauthor{\bsnm{Heger}, \binits{A.}}:
\batitle{{Pulsational pair instability as an explanation for the most luminous supernovae}}.
\bjtitle{Nature}
\bvolume{450},
\bfpage{390}
(\byear{2007})
\end{barticle}
\endbibitem

\bibitem[\protect\citeauthoryear{{Farmer} et~al.}{2019}]{Farmer_2019}
\begin{barticle}
\bauthor{\bsnm{{Farmer}}, \binits{R.}},
\bauthor{\bsnm{{Renzo}}, \binits{M.}},
\bauthor{\bsnm{{de Mink}}, \binits{S.E.}},
\bauthor{\bsnm{{Marchant}}, \binits{P.}},
\bauthor{\bsnm{{Justham}}, \binits{S.}}:
\batitle{{Mind the Gap: The Location of the Lower Edge of the Pair-instability Supernova Black Hole Mass Gap}}.
\bjtitle{Astrophys. J.}
\bvolume{887}(\bissue{1}),
\bfpage{53}
(\byear{2019})
\end{barticle}
\endbibitem

\bibitem[\protect\citeauthoryear{{Abbott} et~al.}{2019}]{2019ApJ...882L..24A}
\begin{barticle}
\bauthor{\bsnm{{Abbott}}, \binits{B.P.}},
\bauthor{\bsnm{{Abbott}}, \binits{R.}},
\bauthor{\bsnm{{Abbott}}, \binits{T.D.}},
\bauthor{\bsnm{{Abraham}}, \binits{S.}},
\bauthor{\bsnm{{Acernese}}, \binits{F.}},
\bauthor{\bsnm{{Ackley}}, \binits{K.}},
\bauthor{\bsnm{{Adams}}, \binits{C.}},
\bauthor{\bsnm{{Adhikari}}, \binits{R.X.}},
\bauthor{\bsnm{{Adya}}, \binits{V.B.}},
\bauthor{\bsnm{{Affeldt}}, \binits{C.}},
\bauthor{\bsnm{al.}}:
\batitle{{Binary Black Hole Population Properties Inferred from the First and Second Observing Runs of Advanced LIGO and Advanced Virgo}}.
\bjtitle{Astrophys. J. Lett.}
\bvolume{882}(\bissue{2}),
\bfpage{24}
(\byear{2019})
\end{barticle}
\endbibitem

\bibitem[\protect\citeauthoryear{Abbott et~al.}{2021}]{gwtc2_pop}
\begin{barticle}
\bauthor{\bsnm{Abbott}, \binits{R.}},
\bauthor{\bsnm{Abbott}, \binits{T.D.}},
\bauthor{\bsnm{Abraham}, \binits{S.}},
\bauthor{\bsnm{Acernese}, \binits{F.}},
\bauthor{\bsnm{al.}}:
\batitle{Population properties of compact objects from the second {LIGO}{\textendash}virgo gravitational-wave transient catalog}.
\bjtitle{The Astrophysical Journal Letters}
\bvolume{913},
\bfpage{7}
(\byear{2021})
\end{barticle}
\endbibitem

\bibitem[\protect\citeauthoryear{Schulze et~al.}{2024}]{Schulze:2023vik}
\begin{barticle}
\bauthor{\bsnm{Schulze}, \binits{S.}}, \betal:
\batitle{{1100 days in the life of the supernova 2018ibb - The best pair-instability supernova candidate, to date}}.
\bjtitle{Astron. Astrophys.}
\bvolume{683},
\bfpage{223}
(\byear{2024})
\end{barticle}
\endbibitem

\bibitem[\protect\citeauthoryear{Angus et~al.}{2024}]{angus}
\begin{barticle}
\bauthor{\bsnm{Angus}, \binits{C.R.}}, \betal:
\batitle{Double acct: a distinct double-peaked supernova matching pulsational pair-instability models}.
\bjtitle{Astrophys. J. Lett.}
\bvolume{977},
\bfpage{41}
(\byear{2024})
\end{barticle}
\endbibitem

\bibitem[\protect\citeauthoryear{Aasi et~al.}{2015}]{aLIGO}
\begin{barticle}
\bauthor{\bsnm{Aasi}, \binits{J.}}, \betal:
\batitle{{Advanced LIGO}}.
\bjtitle{Class. Quant. Grav.}
\bvolume{32},
\bfpage{074001}
(\byear{2015})
\end{barticle}
\endbibitem

\bibitem[\protect\citeauthoryear{Acernese et~al.}{2015}]{aVIRGO}
\begin{barticle}
\bauthor{\bsnm{Acernese}, \binits{F.}}, \betal:
\batitle{{Advanced Virgo: a second-generation interferometric gravitational wave detector}}.
\bjtitle{Class. Quant. Grav.}
\bvolume{32},
\bfpage{024001}
(\byear{2015})
\end{barticle}
\endbibitem

\bibitem[\protect\citeauthoryear{Akutsu et~al.}{2021}]{KAGRA:2020tym}
\begin{barticle}
\bauthor{\bsnm{Akutsu}, \binits{T.}}, \betal:
\batitle{{Overview of KAGRA: Detector design and construction history}}.
\bjtitle{PTEP}
\bvolume{2021}(\bissue{5}),
\bfpage{05}--\blpage{101}
(\byear{2021})
\end{barticle}
\endbibitem

\bibitem[\protect\citeauthoryear{Farmer et~al.}{2020}]{Farmer:2020xne}
\begin{barticle}
\bauthor{\bsnm{Farmer}, \binits{R.}},
\bauthor{\bsnm{Renzo}, \binits{M.}},
\bauthor{\bsnm{Mink}, \binits{S.}},
\bauthor{\bsnm{Fishbach}, \binits{M.}},
\bauthor{\bsnm{Justham}, \binits{S.}}:
\batitle{{Constraints from gravitational wave detections of binary black hole mergers on the $^{12}\rm{C}\left(\alpha,\gamma\right)^{16}\!\rm{O}$ rate}}.
\bjtitle{Astrophys. J. Lett.}
\bvolume{902}(\bissue{2}),
\bfpage{36}
(\byear{2020})
\end{barticle}
\endbibitem

\bibitem[\protect\citeauthoryear{Abbott et~al.}{2023}]{gwtc3_pop}
\begin{barticle}
\bauthor{\bsnm{Abbott}, \binits{R.}}, \betal:
\batitle{{The population of merging compact binaries inferred using gravitational waves through GWTC-3}}.
\bjtitle{Phys. Rev. X}
\bvolume{13},
\bfpage{011048}
(\byear{2023})
\end{barticle}
\endbibitem

\bibitem[\protect\citeauthoryear{Edelman et~al.}{2021}]{Edelman:2021fik}
\begin{barticle}
\bauthor{\bsnm{Edelman}, \binits{B.}},
\bauthor{\bsnm{Doctor}, \binits{Z.}},
\bauthor{\bsnm{Farr}, \binits{B.}}:
\batitle{{Poking Holes: Looking for Gaps in LIGO/Virgo\textquoteright{}s Black Hole Population}}.
\bjtitle{Astrophys. J. Lett.}
\bvolume{913}(\bissue{2}),
\bfpage{23}
(\byear{2021})
\end{barticle}
\endbibitem

\bibitem[\protect\citeauthoryear{Mould et~al.}{2022}]{Mould:2022ccw}
\begin{barticle}
\bauthor{\bsnm{Mould}, \binits{M.}},
\bauthor{\bsnm{Gerosa}, \binits{D.}},
\bauthor{\bsnm{Taylor}, \binits{S.R.}}:
\batitle{{Deep learning and Bayesian inference of gravitational-wave populations: Hierarchical black-hole mergers}}.
\bjtitle{Phys. Rev. D}
\bvolume{106}(\bissue{10}),
\bfpage{103013}
(\byear{2022})
\end{barticle}
\endbibitem

\bibitem[\protect\citeauthoryear{Wang et~al.}{2022}]{Wang:2022gnx}
\begin{barticle}
\bauthor{\bsnm{Wang}, \binits{Y.-Z.}},
\bauthor{\bsnm{Li}, \binits{Y.-J.}},
\bauthor{\bsnm{Vink}, \binits{J.S.}},
\bauthor{\bsnm{Fan}, \binits{Y.-Z.}},
\bauthor{\bsnm{Tang}, \binits{S.-P.}},
\bauthor{\bsnm{Qin}, \binits{Y.}},
\bauthor{\bsnm{Wei}, \binits{D.-M.}}:
\batitle{{Potential Subpopulations and Assembling Tendency of the Merging Black Holes}}.
\bjtitle{Astrophys. J. Lett.}
\bvolume{941}(\bissue{2}),
\bfpage{39}
(\byear{2022})
\end{barticle}
\endbibitem

\bibitem[\protect\citeauthoryear{Li et~al.}{2024}]{Li_2023yyt}
\begin{barticle}
\bauthor{\bsnm{Li}, \binits{Y.-J.}},
\bauthor{\bsnm{Wang}, \binits{Y.-Z.}},
\bauthor{\bsnm{Tang}, \binits{S.-P.}},
\bauthor{\bsnm{Fan}, \binits{Y.-Z.}}:
\batitle{{Resolving the Stellar-Collapse and Hierarchical-Merger Origins of the Coalescing Black Holes}}.
\bjtitle{Phys. Rev. Lett.}
\bvolume{133}(\bissue{5}),
\bfpage{051401}
(\byear{2024})
\end{barticle}
\endbibitem

\bibitem[\protect\citeauthoryear{Antonini et~al.}{2025}]{antonini2024}
\begin{barticle}
\bauthor{\bsnm{Antonini}, \binits{F.}},
\bauthor{\bsnm{Romero-Shaw}, \binits{I.M.}},
\bauthor{\bsnm{Callister}, \binits{T.}}:
\batitle{Star cluster population of high mass black hole mergers in gravitational wave data}.
\bjtitle{Phys. Rev. Lett.}
\bvolume{134},
\bfpage{011401}
(\byear{2025})
\end{barticle}
\endbibitem

\bibitem[\protect\citeauthoryear{Talbot and Thrane}{2018}]{mass}
\begin{barticle}
\bauthor{\bsnm{Talbot}, \binits{C.}},
\bauthor{\bsnm{Thrane}, \binits{E.}}:
\batitle{Measuring the binary black hole mass spectrum with an astrophysically motivated parameterization}.
\bjtitle{Astrophys. J.}
\bvolume{856},
\bfpage{173}
(\byear{2018})
\end{barticle}
\endbibitem

\bibitem[\protect\citeauthoryear{Golomb et~al.}{2024}]{Golomb:2023vxm}
\begin{barticle}
\bauthor{\bsnm{Golomb}, \binits{J.}},
\bauthor{\bsnm{Isi}, \binits{M.}},
\bauthor{\bsnm{Farr}, \binits{W.M.}}:
\batitle{{Physical Models for the Astrophysical Population of Black Holes: Application to the Bump in the Mass Distribution of Gravitational-wave Sources}}.
\bjtitle{Astrophys. J.}
\bvolume{976}(\bissue{1}),
\bfpage{121}
(\byear{2024})
\end{barticle}
\endbibitem

\bibitem[\protect\citeauthoryear{Stevenson et~al.}{2019}]{Stevenson:2019rcw}
\begin{botherref}
\oauthor{\bsnm{Stevenson}, \binits{S.}},
\oauthor{\bsnm{Sampson}, \binits{M.}},
\oauthor{\bsnm{Powell}, \binits{J.}},
\oauthor{\bsnm{Vigna-G\'omez}, \binits{A.}},
\oauthor{\bsnm{Neijssel}, \binits{C.J.}},
\oauthor{\bsnm{Sz\'ecsi}, \binits{D.}},
\oauthor{\bsnm{Mandel}, \binits{I.}}:
{The impact of pair-instability mass loss on the binary black hole mass distribution}
(2019)
\end{botherref}
\endbibitem

\bibitem[\protect\citeauthoryear{Croon and Sakstein}{2023}]{Croon}
\begin{botherref}
\oauthor{\bsnm{Croon}, \binits{D.}},
\oauthor{\bsnm{Sakstein}, \binits{J.}}:
{Prediction of Multiple Features in the Black Hole Mass Function due to Pulsational Pair-Instability Supernovae}
(2023)
{\href{https://arxiv.org/abs/2312.13459}{{arXiv:2312.13459}}}
\end{botherref}
\endbibitem

\bibitem[\protect\citeauthoryear{Gerosa and Fishbach}{2021}]{GerosaFishbach}
\begin{botherref}
\oauthor{\bsnm{Gerosa}, \binits{D.}},
\oauthor{\bsnm{Fishbach}, \binits{M.}}:
Hierarchical mergers of stellar-mass black holes and their gravitational-wave signatures.
Nature
\textbf{5}
(2021)
\end{botherref}
\endbibitem

\bibitem[\protect\citeauthoryear{Di~Carlo et~al.}{2019}]{DiCarlo:2019pmf}
\begin{barticle}
\bauthor{\bsnm{Di~Carlo}, \binits{U.N.}},
\bauthor{\bsnm{Giacobbo}, \binits{N.}},
\bauthor{\bsnm{Mapelli}, \binits{M.}},
\bauthor{\bsnm{Pasquato}, \binits{M.}},
\bauthor{\bsnm{Spera}, \binits{M.}},
\bauthor{\bsnm{Wang}, \binits{L.}},
\bauthor{\bsnm{Haardt}, \binits{F.}}:
\batitle{{Merging black holes in young star clusters}}.
\bjtitle{Mon. Not. Roy. Astron. Soc.}
\bvolume{487}(\bissue{2}),
\bfpage{2947}--\blpage{2960}
(\byear{2019})
\end{barticle}
\endbibitem

\bibitem[\protect\citeauthoryear{Renzo et~al.}{2020}]{Renzo:2020smh}
\begin{barticle}
\bauthor{\bsnm{Renzo}, \binits{M.}},
\bauthor{\bsnm{Cantiello}, \binits{M.}},
\bauthor{\bsnm{Metzger}, \binits{B.D.}},
\bauthor{\bsnm{Jiang}, \binits{Y.-F.}}:
\batitle{{The Stellar Merger Scenario for Black Holes in the Pair-instability Gap}}.
\bjtitle{Astrophys. J. Lett.}
\bvolume{904}(\bissue{2}),
\bfpage{13}
(\byear{2020})
\end{barticle}
\endbibitem

\bibitem[\protect\citeauthoryear{Siegel et~al.}{2022}]{Siegel:2021ptt}
\begin{barticle}
\bauthor{\bsnm{Siegel}, \binits{D.M.}},
\bauthor{\bsnm{Agarwal}, \binits{A.}},
\bauthor{\bsnm{Barnes}, \binits{J.}},
\bauthor{\bsnm{Metzger}, \binits{B.D.}},
\bauthor{\bsnm{Renzo}, \binits{M.}},
\bauthor{\bsnm{Villar}, \binits{V.A.}}:
\batitle{{{\textquotedblleft}Super-kilonovae{\textquotedblright} from Massive Collapsars as Signatures of Black Hole Birth in the Pair-instability Mass Gap}}.
\bjtitle{Astrophys. J.}
\bvolume{941}(\bissue{1}),
\bfpage{100}
(\byear{2022})
\end{barticle}
\endbibitem

\bibitem[\protect\citeauthoryear{{McKernan} et~al.}{2012}]{McKernan_2012}
\begin{barticle}
\bauthor{\bsnm{{McKernan}}, \binits{B.}},
\bauthor{\bsnm{{Ford}}, \binits{K.E.S.}},
\bauthor{\bsnm{{Lyra}}, \binits{W.}},
\bauthor{\bsnm{{Perets}}, \binits{H.B.}}:
\batitle{{Intermediate mass black holes in AGN discs - I. Production and growth}}.
\bjtitle{Mon. Not. R. Ast. Soc.}
\bvolume{425}(\bissue{1}),
\bfpage{460}--\blpage{469}
(\byear{2012})
\end{barticle}
\endbibitem

\bibitem[\protect\citeauthoryear{Abac et~al.}{2025}]{GWTC-4_result}
\begin{botherref}
\oauthor{\bsnm{Abac}, \binits{A.G.}}, et al.:
{GWTC-4.0: Updating the Gravitational-Wave Transient Catalog with Observations from the First Part of the Fourth LIGO-Virgo-KAGRA Observing Run}
(2025)
{\href{https://arxiv.org/abs/2508.18082}{{arXiv:2508.18082}}}
{[gr-qc]}
\end{botherref}
\endbibitem

\bibitem[\protect\citeauthoryear{Woosley}{2017}]{Woosley:2016hmi}
\begin{barticle}
\bauthor{\bsnm{Woosley}, \binits{S.E.}}:
\batitle{{Pulsational Pair-Instability Supernovae}}.
\bjtitle{Astrophys. J.}
\bvolume{836}(\bissue{2}),
\bfpage{244}
(\byear{2017})
\end{barticle}
\endbibitem

\bibitem[\protect\citeauthoryear{Abac et~al.}{2025}]{GW231123}
\begin{botherref}
\oauthor{\bsnm{Abac}, \binits{A.G.}}, et al.:
{GW231123: a Binary Black Hole Merger with Total Mass 190-265 $M_{\odot}$}
(2025)
{\href{https://arxiv.org/abs/2507.08219}{{arXiv:2507.08219}}}
{[astro-ph.HE]}
\end{botherref}
\endbibitem

\bibitem[\protect\citeauthoryear{Gottlieb et~al.}{2025}]{Gottlieb:2025ugy}
\begin{botherref}
\oauthor{\bsnm{Gottlieb}, \binits{O.}},
\oauthor{\bsnm{Metzger}, \binits{B.D.}},
\oauthor{\bsnm{Issa}, \binits{D.}},
\oauthor{\bsnm{Li}, \binits{S.E.}},
\oauthor{\bsnm{Renzo}, \binits{M.}},
\oauthor{\bsnm{Isi}, \binits{M.}}:
{Spinning into the Gap: Direct-Horizon Collapse as the Origin of GW231123 from End-to-End GRMHD Simulations}
(2025)
{\href{https://arxiv.org/abs/2508.15887}{{arXiv:2508.15887}}}
{[astro-ph.HE]}
\end{botherref}
\endbibitem

\bibitem[\protect\citeauthoryear{Payne and Thrane}{2023}]{PhysRevResearch.5.023013}
\begin{barticle}
\bauthor{\bsnm{Payne}, \binits{E.}},
\bauthor{\bsnm{Thrane}, \binits{E.}}:
\batitle{Model exploration in gravitational-wave astronomy with the maximum population likelihood}.
\bjtitle{Phys. Rev. Res.}
\bvolume{5},
\bfpage{023013}
(\byear{2023})
\end{barticle}
\endbibitem

\bibitem[\protect\citeauthoryear{Fitchett}{1983}]{Fitchett_1983}
\begin{barticle}
\bauthor{\bsnm{Fitchett}, \binits{M.J.}}:
\batitle{The influence of gravitational wave momentum losses on the centre of mass motion of a newtonian binary system}.
\bjtitle{Monthly Notices of the Royal Astronomical Society}
\bvolume{203}(\bissue{4}),
\bfpage{1049}--\blpage{1062}
(\byear{1983})
\end{barticle}
\endbibitem

\bibitem[\protect\citeauthoryear{Gerosa et~al.}{2018}]{Gerosa:2018qay}
\begin{barticle}
\bauthor{\bsnm{Gerosa}, \binits{D.}},
\bauthor{\bsnm{H{\'e}bert}, \binits{F.}},
\bauthor{\bsnm{Stein}, \binits{L.C.}}:
\batitle{{Black-hole kicks from numerical-relativity surrogate models}}.
\bjtitle{Phys. Rev. D}
\bvolume{97}(\bissue{10}),
\bfpage{104049}
(\byear{2018})
\end{barticle}
\endbibitem

\bibitem[\protect\citeauthoryear{Antonini and Rasio}{2016}]{Antonini:2016gqe}
\begin{barticle}
\bauthor{\bsnm{Antonini}, \binits{F.}},
\bauthor{\bsnm{Rasio}, \binits{F.A.}}:
\batitle{{Merging black hole binaries in galactic nuclei: implications for advanced-LIGO detections}}.
\bjtitle{Astrophys. J.}
\bvolume{831}(\bissue{2}),
\bfpage{187}
(\byear{2016})
\end{barticle}
\endbibitem

\bibitem[\protect\citeauthoryear{Rodriguez et~al.}{2019}]{Rodriguez:2019huv}
\begin{barticle}
\bauthor{\bsnm{Rodriguez}, \binits{C.L.}},
\bauthor{\bsnm{Zevin}, \binits{M.}},
\bauthor{\bsnm{Amaro-Seoane}, \binits{P.}},
\bauthor{\bsnm{Chatterjee}, \binits{S.}},
\bauthor{\bsnm{Kremer}, \binits{K.}},
\bauthor{\bsnm{Rasio}, \binits{F.A.}},
\bauthor{\bsnm{Ye}, \binits{C.S.}}:
\batitle{{Black holes: The next generation\textemdash{}repeated mergers in dense star clusters and their gravitational-wave properties}}.
\bjtitle{Phys. Rev. D}
\bvolume{100}(\bissue{4}),
\bfpage{043027}
(\byear{2019})
\end{barticle}
\endbibitem

\bibitem[\protect\citeauthoryear{Abac et~al.}{2025}]{GWTC_4_population}
\begin{botherref}
\oauthor{\bsnm{Abac}, \binits{A.G.}}, et al.:
{GWTC-4.0: Population Properties of Merging Compact Binaries}
(2025)
{\href{https://arxiv.org/abs/2508.18083}{{arXiv:2508.18083}}}
{[astro-ph.HE]}
\end{botherref}
\endbibitem

\bibitem[\protect\citeauthoryear{Pierra et~al.}{2024}]{Pierra_2024}
\begin{barticle}
\bauthor{\bsnm{Pierra}, \binits{G.}},
\bauthor{\bsnm{Mastrogiovanni}, \binits{S.}},
\bauthor{\bsnm{Perriès}, \binits{S.}}:
\batitle{The spin magnitude of stellar-mass black holes evolves with the mass}.
\bjtitle{Astron. Astrophys.}
\bvolume{692},
\bfpage{80}
(\byear{2024})
\end{barticle}
\endbibitem

\bibitem[\protect\citeauthoryear{Antonini et~al.}{2025}]{antonini2025_PISN}
\begin{botherref}
\oauthor{\bsnm{Antonini}, \binits{F.}},
\oauthor{\bsnm{Romero-Shaw}, \binits{I.}},
\oauthor{\bsnm{Callister}, \binits{T.}},
\oauthor{\bsnm{Dosopoulou}, \binits{F.}},
\oauthor{\bsnm{Chattopadhyay}, \binits{D.}},
\oauthor{\bsnm{Gieles}, \binits{M.}},
\oauthor{\bsnm{Mapelli}, \binits{M.}}:
Gravitational waves reveal the pair-instability mass gap and constrain nuclear burning in massive stars
(2025)
{\href{https://arxiv.org/abs/2509.04637}{{arXiv:2509.04637}}}
{[astro-ph.HE]}
\end{botherref}
\endbibitem

\bibitem[\protect\citeauthoryear{McKernan and Ford}{2024}]{McKernan:2023xio}
\begin{barticle}
\bauthor{\bsnm{McKernan}, \binits{B.}},
\bauthor{\bsnm{Ford}, \binits{K.E.S.}}:
\batitle{{Constraining the LVK AGN channel with black hole spins}}.
\bjtitle{Mon. Not. Roy. Astron. Soc.}
\bvolume{531}(\bissue{3}),
\bfpage{3479}--\blpage{3485}
(\byear{2024})
\doiurl{10.1093/mnras/stae1351}
\end{barticle}
\endbibitem

\bibitem[\protect\citeauthoryear{Fishbach and Holz}{2020}]{Fishbach_GW190521}
\begin{barticle}
\bauthor{\bsnm{Fishbach}, \binits{M.}},
\bauthor{\bsnm{Holz}, \binits{D.E.}}:
\batitle{Minding the gap: Gw190521 as a straddling binary}.
\bjtitle{The Astrophysical Journal Letters}
\bvolume{904}(\bissue{2}),
\bfpage{26}
(\byear{2020})
\end{barticle}
\endbibitem

\bibitem[\protect\citeauthoryear{deBoer et~al.}{2017}]{deBoer_2017}
\begin{barticle}
\bauthor{\bsnm{deBoer}, \binits{R.J.}},
\bauthor{\bsnm{G\"orres}, \binits{J.}},
\bauthor{\bsnm{Wiescher}, \binits{M.}},
\bauthor{\bsnm{Azuma}, \binits{R.E.}},
\bauthor{\bsnm{Best}, \binits{A.}},
\bauthor{\bsnm{Brune}, \binits{C.R.}},
\bauthor{\bsnm{Fields}, \binits{C.E.}},
\bauthor{\bsnm{Jones}, \binits{S.}},
\bauthor{\bsnm{Pignatari}, \binits{M.}},
\bauthor{\bsnm{Sayre}, \binits{D.}},
\bauthor{\bsnm{Smith}, \binits{K.}},
\bauthor{\bsnm{Timmes}, \binits{F.X.}},
\bauthor{\bsnm{Uberseder}, \binits{E.}}:
\batitle{The $^{12}\mathbf{C}(\ensuremath{\alpha},\ensuremath{\gamma})^{16}\mathbf{O}$ reaction and its implications for stellar helium burning}.
\bjtitle{Rev. Mod. Phys.}
\bvolume{89},
\bfpage{035007}
(\byear{2017})
\end{barticle}
\endbibitem

\bibitem[\protect\citeauthoryear{{Buchmann}}{1996}]{Buchmann_1996}
\begin{barticle}
\bauthor{\bsnm{{Buchmann}}, \binits{L.}}:
\batitle{{New Stellar Reaction Rate for 12C( alpha , gamma ) 16O}}.
\bjtitle{Astrophys. J. Lett.}
\bvolume{468},
\bfpage{127}
(\byear{1996})
\doiurl{10.1086/310240}
\end{barticle}
\endbibitem

\bibitem[\protect\citeauthoryear{{Kunz} et~al.}{2002}]{Kunz_2002}
\begin{barticle}
\bauthor{\bsnm{{Kunz}}, \binits{R.}},
\bauthor{\bsnm{{Fey}}, \binits{M.}},
\bauthor{\bsnm{{Jaeger}}, \binits{M.}},
\bauthor{\bsnm{{Mayer}}, \binits{A.}},
\bauthor{\bsnm{{Hammer}}, \binits{J.W.}},
\bauthor{\bsnm{{Staudt}}, \binits{G.}},
\bauthor{\bsnm{{Harissopulos}}, \binits{S.}},
\bauthor{\bsnm{{Paradellis}}, \binits{T.}}:
\batitle{{Astrophysical Reaction Rate of $^{12}$C({\ensuremath{\alpha}}, {\ensuremath{\gamma}})$^{16}$O}}.
\bjtitle{Astrophys. J.}
\bvolume{567}(\bissue{1}),
\bfpage{643}--\blpage{650}
(\byear{2002})
\doiurl{10.1086/338384}
\end{barticle}
\endbibitem

\bibitem[\protect\citeauthoryear{{Shen} et~al.}{2023}]{2023ApJ...945...41S}
\begin{barticle}
\bauthor{\bsnm{{Shen}}, \binits{Y.}}, \betal:
\batitle{{New Determination of the $^{12}$C({\ensuremath{\alpha}}, {\ensuremath{\gamma}})$^{16}$O Reaction Rate and Its Impact on the Black-hole Mass Gap}}.
\bjtitle{Astrophys. J.}
\bvolume{945}(\bissue{1}),
\bfpage{41}
(\byear{2023})
\end{barticle}
\endbibitem

\bibitem[\protect\citeauthoryear{Farr et~al.}{2019}]{Farr_2019}
\begin{barticle}
\bauthor{\bsnm{Farr}, \binits{W.M.}},
\bauthor{\bsnm{Fishbach}, \binits{M.}},
\bauthor{\bsnm{Ye}, \binits{J.}},
\bauthor{\bsnm{Holz}, \binits{D.E.}}:
\batitle{A future percent-level measurement of the hubble expansion at redshift 0.8 with advanced ligo}.
\bjtitle{The Astrophysical Journal Letters}
\bvolume{883},
\bfpage{42}
(\byear{2019})
\end{barticle}
\endbibitem

\end{thebibliography}

\begin{thebibliography}{49}
\makeatletter
\addtocounter{NAT@ctr}{49}
\makeatother
\ifx \bisbn   \undefined \def \bisbn  #1{ISBN #1}\fi
\ifx \binits  \undefined \def \binits#1{#1}\fi
\ifx \bauthor  \undefined \def \bauthor#1{#1}\fi
\ifx \batitle  \undefined \def \batitle#1{#1}\fi
\ifx \bjtitle  \undefined \def \bjtitle#1{#1}\fi
\ifx \bvolume  \undefined \def \bvolume#1{\textbf{#1}}\fi
\ifx \byear  \undefined \def \byear#1{#1}\fi
\ifx \bissue  \undefined \def \bissue#1{#1}\fi
\ifx \bfpage  \undefined \def \bfpage#1{#1}\fi
\ifx \blpage  \undefined \def \blpage #1{#1}\fi
\ifx \burl  \undefined \def \burl#1{\textsf{#1}}\fi
\ifx \doiurl  \undefined \def \doiurl#1{\url{https://doi.org/#1}}\fi
\ifx \betal  \undefined \def \betal{\textit{et al.}}\fi
\ifx \binstitute  \undefined \def \binstitute#1{#1}\fi
\ifx \binstitutionaled  \undefined \def \binstitutionaled#1{#1}\fi
\ifx \bctitle  \undefined \def \bctitle#1{#1}\fi
\ifx \beditor  \undefined \def \beditor#1{#1}\fi
\ifx \bpublisher  \undefined \def \bpublisher#1{#1}\fi
\ifx \bbtitle  \undefined \def \bbtitle#1{#1}\fi
\ifx \bedition  \undefined \def \bedition#1{#1}\fi
\ifx \bseriesno  \undefined \def \bseriesno#1{#1}\fi
\ifx \blocation  \undefined \def \blocation#1{#1}\fi
\ifx \bsertitle  \undefined \def \bsertitle#1{#1}\fi
\ifx \bsnm \undefined \def \bsnm#1{#1}\fi
\ifx \bsuffix \undefined \def \bsuffix#1{#1}\fi
\ifx \bparticle \undefined \def \bparticle#1{#1}\fi
\ifx \barticle \undefined \def \barticle#1{#1}\fi
\bibcommenthead
\ifx \bconfdate \undefined \def \bconfdate #1{#1}\fi
\ifx \botherref \undefined \def \botherref #1{#1}\fi
\ifx \url \undefined \def \url#1{\textsf{#1}}\fi
\ifx \bchapter \undefined \def \bchapter#1{#1}\fi
\ifx \bbook \undefined \def \bbook#1{#1}\fi
\ifx \bcomment \undefined \def \bcomment#1{#1}\fi
\ifx \oauthor \undefined \def \oauthor#1{#1}\fi
\ifx \citeauthoryear \undefined \def \citeauthoryear#1{#1}\fi
\ifx \endbibitem  \undefined \def \endbibitem {}\fi
\ifx \bconflocation  \undefined \def \bconflocation#1{#1}\fi
\ifx \arxivurl  \undefined \def \arxivurl#1{\textsf{#1}}\fi
\csname PreBibitemsHook\endcsname

\bibitem[\protect\citeauthoryear{Talbot et~al.}{2025}]{gwpopulation}
\begin{barticle}
\bauthor{\bsnm{Talbot}, \binits{C.}},
\bauthor{\bsnm{Farah}, \binits{A.}},
\bauthor{\bsnm{Galaudage}, \binits{S.}},
\bauthor{\bsnm{Golomb}, \binits{J.}},
\bauthor{\bsnm{Tong}, \binits{H.}}:
\batitle{{GWPopulation: Hardware agnostic population inference for compact binaries and beyond}}.
\bjtitle{J. Open Source Softw.}
\bvolume{10}(\bissue{109}),
\bfpage{7753}
(\byear{2025})
\end{barticle}
\endbibitem

\bibitem[\protect\citeauthoryear{Fishbach et~al.}{2018}]{Fishbach:2018edt}
\begin{barticle}
\bauthor{\bsnm{Fishbach}, \binits{M.}},
\bauthor{\bsnm{Holz}, \binits{D.E.}},
\bauthor{\bsnm{Farr}, \binits{W.M.}}:
\batitle{{Does the Black Hole Merger Rate Evolve with Redshift?}}
\bjtitle{Astrophys. J. Lett.}
\bvolume{863}(\bissue{2}),
\bfpage{41}
(\byear{2018})
\end{barticle}
\endbibitem

\bibitem[\protect\citeauthoryear{Thrane and Talbot}{2019}]{2019_Bayesian}
\begin{botherref}
\oauthor{\bsnm{Thrane}, \binits{E.}},
\oauthor{\bsnm{Talbot}, \binits{C.}}:
An introduction to bayesian inference in gravitational-wave astronomy: Parameter estimation, model selection, and hierarchical models.
Publications of the Astronomical Society of Australia
\textbf{36}
(2019)
\end{botherref}
\endbibitem

\bibitem[\protect\citeauthoryear{Mandel et~al.}{2019}]{Mandel_2019}
\begin{barticle}
\bauthor{\bsnm{Mandel}, \binits{I.}},
\bauthor{\bsnm{Farr}, \binits{W.M.}},
\bauthor{\bsnm{Gair}, \binits{J.R.}}:
\batitle{Extracting distribution parameters from multiple uncertain observations with selection biases}.
\bjtitle{Monthly Notices of the Royal Astronomical Society}
\bvolume{486}(\bissue{1}),
\bfpage{1086}--\blpage{1093}
(\byear{2019})
\end{barticle}
\endbibitem

\bibitem[\protect\citeauthoryear{Vitale et~al.}{2020}]{Vitale:2020aaz}
\begin{botherref}
\oauthor{\bsnm{Vitale}, \binits{S.}},
\oauthor{\bsnm{Gerosa}, \binits{D.}},
\oauthor{\bsnm{Farr}, \binits{W.M.}},
\oauthor{\bsnm{Taylor}, \binits{S.R.}}:
{Inferring the properties of a population of compact binaries in presence of selection effects}
(2020)
{\href{https://arxiv.org/abs/2007.05579}{{arXiv:2007.05579}}}
{[astro-ph.IM]}
\end{botherref}
\endbibitem

\bibitem[\protect\citeauthoryear{Abac et~al.}{2025}]{GWTC-4_method}
\begin{botherref}
\oauthor{\bsnm{Abac}, \binits{A.G.}}, et al.:
{GWTC-4.0: Methods for Identifying and Characterizing Gravitational-wave Transients}
(2025)
{\href{https://arxiv.org/abs/2508.18081}{{arXiv:2508.18081}}}
{[gr-qc]}
\end{botherref}
\endbibitem

\bibitem[\protect\citeauthoryear{{Buikema} et~al.}{2020}]{2020PhRvD.102f2003B}
\begin{barticle}
\bauthor{\bsnm{{Buikema}}, \binits{A.}}, \betal:
\batitle{{Sensitivity and performance of the Advanced LIGO detectors in the third observing run}}.
\bjtitle{Phys. Rev. D}
\bvolume{102}(\bissue{6}),
\bfpage{062003}
(\byear{2020})
\end{barticle}
\endbibitem

\bibitem[\protect\citeauthoryear{{Ganapathy} et~al.}{2023}]{2023PhRvX..13d1021G}
\begin{barticle}
\bauthor{\bsnm{{Ganapathy}}, \binits{D.}}, \betal:
\batitle{{Broadband Quantum Enhancement of the LIGO Detectors with Frequency-Dependent Squeezing}}.
\bjtitle{Physical Review X}
\bvolume{13}(\bissue{4}),
\bfpage{041021}
(\byear{2023})
\end{barticle}
\endbibitem

\bibitem[\protect\citeauthoryear{Jia et~al.}{2024}]{LIGOScientific:2024elc}
\begin{barticle}
\bauthor{\bsnm{Jia}, \binits{W.}}, \betal:
\batitle{{Squeezing the quantum noise of a gravitational-wave detector below the standard quantum limit}}.
\bjtitle{Science}
\bvolume{385}(\bissue{6715}),
\bfpage{1318}
(\byear{2024})
\end{barticle}
\endbibitem

\bibitem[\protect\citeauthoryear{Capote et~al.}{2025}]{Capote:2024rmo}
\begin{barticle}
\bauthor{\bsnm{Capote}, \binits{E.}}, \betal:
\batitle{{Advanced LIGO detector performance in the fourth observing run}}.
\bjtitle{Phys. Rev. D}
\bvolume{111}(\bissue{6}),
\bfpage{062002}
(\byear{2025})
\end{barticle}
\endbibitem

\bibitem[\protect\citeauthoryear{Soni et~al.}{2025}]{LIGO:2024kkz}
\begin{barticle}
\bauthor{\bsnm{Soni}, \binits{S.}}, \betal:
\batitle{{LIGO Detector Characterization in the first half of the fourth Observing run}}.
\bjtitle{Class. Quant. Grav.}
\bvolume{42}(\bissue{8}),
\bfpage{085016}
(\byear{2025})
\end{barticle}
\endbibitem

\bibitem[\protect\citeauthoryear{Tiwari}{2018}]{Tiwari:2017ndi}
\begin{barticle}
\bauthor{\bsnm{Tiwari}, \binits{V.}}:
\batitle{{Estimation of the Sensitive Volume for Gravitational-wave Source Populations Using Weighted Monte Carlo Integration}}.
\bjtitle{Class. Quant. Grav.}
\bvolume{35}(\bissue{14}),
\bfpage{145009}
(\byear{2018})
\end{barticle}
\endbibitem

\bibitem[\protect\citeauthoryear{{Farr}}{2019}]{2019_Farr_selection}
\begin{barticle}
\bauthor{\bsnm{{Farr}}, \binits{W.M.}}:
\batitle{{Accuracy Requirements for Empirically Measured Selection Functions}}.
\bjtitle{Research Notes of the American Astronomical Society}
\bvolume{3}(\bissue{5}),
\bfpage{66}
(\byear{2019})
\end{barticle}
\endbibitem

\bibitem[\protect\citeauthoryear{Essick and Farr}{2022}]{Essick:2022ojx}
\begin{botherref}
\oauthor{\bsnm{Essick}, \binits{R.}},
\oauthor{\bsnm{Farr}, \binits{W.}}:
{Precision Requirements for Monte Carlo Sums within Hierarchical Bayesian Inference}
(2022)
{[astro-ph.IM]}.
arxiv/2204.00461
\end{botherref}
\endbibitem

\bibitem[\protect\citeauthoryear{Essick et~al.}{2025}]{Essick:2025zed}
\begin{botherref}
\oauthor{\bsnm{Essick}, \binits{R.}}, et al.:
{Compact Binary Coalescence Sensitivity Estimates with Injection Campaigns during the LIGO-Virgo-KAGRA Collaborations' Fourth Observing Run}
(2025)
{\href{https://arxiv.org/abs/2508.10638}{{arXiv:2508.10638}}}
{[gr-qc]}
\end{botherref}
\endbibitem

\bibitem[\protect\citeauthoryear{Ashton et~al.}{2019}]{2019_bilby}
\begin{barticle}
\bauthor{\bsnm{Ashton}, \binits{G.}},
\bauthor{\bsnm{Hübner}, \binits{M.}},
\bauthor{\bsnm{Lasky}, \binits{P.D.}}, \betal:
\batitle{Bilby: A user-friendly bayesian inference library for gravitational-wave astronomy}.
\bjtitle{The Astrophysical Journal Supplement Series}
\bvolume{241}(\bissue{2}),
\bfpage{27}
(\byear{2019})
\end{barticle}
\endbibitem

\bibitem[\protect\citeauthoryear{Romero-Shaw et~al.}{2020}]{2020_bilby}
\begin{barticle}
\bauthor{\bsnm{Romero-Shaw}, \binits{I.M.}},
\bauthor{\bsnm{Talbot}, \binits{C.}}, \betal:
\batitle{Bayesian inference for compact binary coalescences with bilby: validation and application to the first ligo–virgo gravitational-wave transient catalogue}.
\bjtitle{Monthly Notices of the Royal Astronomical Society}
\bvolume{499}(\bissue{3}),
\bfpage{3295}--\blpage{3319}
(\byear{2020})
\end{barticle}
\endbibitem

\bibitem[\protect\citeauthoryear{{Speagle}}{2020}]{dynesty}
\begin{barticle}
\bauthor{\bsnm{{Speagle}}, \binits{J.S.}}:
\batitle{{DYNESTY: a dynamic nested sampling package for estimating Bayesian posteriors and evidences}}.
\bjtitle{Mon. Not. R. Ast. Soc.}
\bvolume{493}(\bissue{3}),
\bfpage{3132}--\blpage{3158}
(\byear{2020})
\end{barticle}
\endbibitem

\bibitem[\protect\citeauthoryear{Abbott et~al.}{2024}]{gwtc-2.1}
\begin{barticle}
\bauthor{\bsnm{Abbott}, \binits{R.}}, \betal:
\batitle{{GWTC-2.1: Deep extended catalog of compact binary coalescences observed by LIGO and Virgo during the first half of the third observing run}}.
\bjtitle{Phys. Rev. D}
\bvolume{109}(\bissue{2}),
\bfpage{022001}
(\byear{2024})
\end{barticle}
\endbibitem

\bibitem[\protect\citeauthoryear{Abbott et~al.}{2023}]{gwtc-3}
\begin{barticle}
\bauthor{\bsnm{Abbott}, \binits{R.}}, \betal:
\batitle{{GWTC-3: Compact Binary Coalescences Observed by LIGO and Virgo During the Second Part of the Third Observing Run}}.
\bjtitle{Phys. Rev. X}
\bvolume{13},
\bfpage{041039}
(\byear{2023})
\end{barticle}
\endbibitem

\bibitem[\protect\citeauthoryear{Abbott et~al.}{2020}]{GW190521_implications}
\begin{barticle}
\bauthor{\bsnm{Abbott}, \binits{R.}}, \betal:
\batitle{{Properties and Astrophysical Implications of the $150 M_\odot$ Binary Black Hole Merger GW190521}}.
\bjtitle{Astrophys. J. Lett.}
\bvolume{900},
\bfpage{13}
(\byear{2020})
\end{barticle}
\endbibitem

\bibitem[\protect\citeauthoryear{Woosley}{2019}]{Woosley_2019}
\begin{barticle}
\bauthor{\bsnm{Woosley}, \binits{S.E.}}:
\batitle{The evolution of massive helium stars, including mass loss}.
\bjtitle{The Astrophysical Journal}
\bvolume{878}(\bissue{1}),
\bfpage{49}
(\byear{2019})
\end{barticle}
\endbibitem

\bibitem[\protect\citeauthoryear{Marchant et~al.}{2018}]{Marchant:2018kun}
\begin{botherref}
\oauthor{\bsnm{Marchant}, \binits{P.}},
\oauthor{\bsnm{Renzo}, \binits{M.}},
\oauthor{\bsnm{Farmer}, \binits{R.}},
\oauthor{\bsnm{Pappas}, \binits{K.M.W.}},
\oauthor{\bsnm{Taam}, \binits{R.E.}},
\oauthor{\bsnm{Mink}, \binits{S.}},
\oauthor{\bsnm{Kalogera}, \binits{V.}}:
{Pulsational pair-instability supernovae in very close binaries}
(2018)
\end{botherref}
\endbibitem

\bibitem[\protect\citeauthoryear{Woosley and Heger}{2021}]{Woosley:2021xba}
\begin{barticle}
\bauthor{\bsnm{Woosley}, \binits{S.E.}},
\bauthor{\bsnm{Heger}, \binits{A.}}:
\batitle{{The Pair-Instability Mass Gap for Black Holes}}.
\bjtitle{Astrophys. J. Lett.}
\bvolume{912}(\bissue{2}),
\bfpage{31}
(\byear{2021})
\end{barticle}
\endbibitem

\bibitem[\protect\citeauthoryear{Renzo and Smith}{2024}]{Renzo:2024jhc}
\begin{botherref}
\oauthor{\bsnm{Renzo}, \binits{M.}},
\oauthor{\bsnm{Smith}, \binits{N.}}:
{Pair-instability evolution and explosions in massive stars}
(2024)
{\href{https://arxiv.org/abs/2407.16113}{{arXiv:2407.16113}}}
{[astro-ph.HE]}
\end{botherref}
\endbibitem

\bibitem[\protect\citeauthoryear{Spera and Mapelli}{2017}]{Spera:2017fyx}
\begin{barticle}
\bauthor{\bsnm{Spera}, \binits{M.}},
\bauthor{\bsnm{Mapelli}, \binits{M.}}:
\batitle{{Very massive stars, pair-instability supernovae and intermediate-mass black holes with the SEVN code}}.
\bjtitle{Mon. Not. Roy. Astron. Soc.}
\bvolume{470}(\bissue{4}),
\bfpage{4739}--\blpage{4749}
(\byear{2017})
\end{barticle}
\endbibitem

\bibitem[\protect\citeauthoryear{Romero-Shaw et~al.}{2022}]{wmf}
\begin{barticle}
\bauthor{\bsnm{Romero-Shaw}, \binits{I.M.}},
\bauthor{\bsnm{Thrane}, \binits{E.}},
\bauthor{\bsnm{Lasky}, \binits{P.D.}}:
\batitle{When models fail: an introduction to posterior predictive checks and model misspecification in gravitational-wave astronomy}.
\bjtitle{Pub. Astron. Soc. Aust.}
\bvolume{39},
\bfpage{025}
(\byear{2022})
\end{barticle}
\endbibitem

\bibitem[\protect\citeauthoryear{Miller et~al.}{2024}]{Miller:2024sui}
\begin{barticle}
\bauthor{\bsnm{Miller}, \binits{S.J.}},
\bauthor{\bsnm{Ko}, \binits{Z.}},
\bauthor{\bsnm{Callister}, \binits{T.}},
\bauthor{\bsnm{Chatziioannou}, \binits{K.}}:
\batitle{{Gravitational waves carry information beyond effective spin parameters but it is hard to extract}}.
\bjtitle{Phys. Rev. D}
\bvolume{109}(\bissue{10}),
\bfpage{104036}
(\byear{2024})
\end{barticle}
\endbibitem

\bibitem[\protect\citeauthoryear{Fishbach et~al.}{2020}]{Fishbach:2020ryj}
\begin{barticle}
\bauthor{\bsnm{Fishbach}, \binits{M.}},
\bauthor{\bsnm{Essick}, \binits{R.}},
\bauthor{\bsnm{Holz}, \binits{D.E.}}:
\batitle{{Does Matter Matter? Using the mass distribution to distinguish neutron stars and black holes}}.
\bjtitle{Astrophys. J. Lett.}
\bvolume{899},
\bfpage{8}
(\byear{2020})
\end{barticle}
\endbibitem

\bibitem[\protect\citeauthoryear{Farah et~al.}{2024}]{Farah:2023swu}
\begin{barticle}
\bauthor{\bsnm{Farah}, \binits{A.M.}},
\bauthor{\bsnm{Fishbach}, \binits{M.}},
\bauthor{\bsnm{Holz}, \binits{D.E.}}:
\batitle{{Two of a Kind: Comparing Big and Small Black Holes in Binaries with Gravitational Waves}}.
\bjtitle{Astrophys. J.}
\bvolume{962}(\bissue{1}),
\bfpage{69}
(\byear{2024})
\end{barticle}
\endbibitem

\bibitem[\protect\citeauthoryear{Antonini et~al.}{2025}]{antonini2025}
\begin{botherref}
\oauthor{\bsnm{Antonini}, \binits{F.}},
\oauthor{\bsnm{Callister}, \binits{T.}},
\oauthor{\bsnm{Dosopoulou}, \binits{F.}},
\oauthor{\bsnm{Romero-Shaw}, \binits{I.}},
\oauthor{\bsnm{Chattopadhyay}, \binits{D.}}:
{Inferring the pair-instability mass gap from gravitational wave data using flexible models}
(2025)
{\href{https://arxiv.org/abs/2506.09154}{{arXiv:2506.09154}}}
{[astro-ph.HE]}
\end{botherref}
\endbibitem

\bibitem[\protect\citeauthoryear{Abbott et~al.}{2020}]{GW190521}
\begin{barticle}
\bauthor{\bsnm{Abbott}, \binits{R.}}, \betal:
\batitle{{GW190521: A Binary Black Hole Merger with a Total Mass of $150 M_\odot$}}.
\bjtitle{Phys. Rev. Lett.}
\bvolume{125},
\bfpage{101102}
(\byear{2020})
\end{barticle}
\endbibitem

\bibitem[\protect\citeauthoryear{{Galaudage} et~al.}{2020}]{2020PhRvD.102h3026G}
\begin{barticle}
\bauthor{\bsnm{{Galaudage}}, \binits{S.}},
\bauthor{\bsnm{{Talbot}}, \binits{C.}},
\bauthor{\bsnm{{Thrane}}, \binits{E.}}:
\batitle{{Gravitational-wave inference in the catalog era: Evolving priors and marginal events}}.
\bjtitle{Phys. Rev. D}
\bvolume{102}(\bissue{8}),
\bfpage{083026}
(\byear{2020})
\end{barticle}
\endbibitem

\bibitem[\protect\citeauthoryear{{Fishbach} et~al.}{2020}]{2020ApJ...891L..31F}
\begin{barticle}
\bauthor{\bsnm{{Fishbach}}, \binits{M.}},
\bauthor{\bsnm{{Farr}}, \binits{W.M.}},
\bauthor{\bsnm{{Holz}}, \binits{D.E.}}:
\batitle{{The Most Massive Binary Black Hole Detections and the Identification of Population Outliers}}.
\bjtitle{Astrophys. J. Lett.}
\bvolume{891}(\bissue{2}),
\bfpage{31}
(\byear{2020})
\end{barticle}
\endbibitem

\bibitem[\protect\citeauthoryear{Doctor et~al.}{2020}]{Doctor_2020}
\begin{barticle}
\bauthor{\bsnm{Doctor}, \binits{Z.}},
\bauthor{\bsnm{Wysocki}, \binits{D.}},
\bauthor{\bsnm{O’Shaughnessy}, \binits{R.}},
\bauthor{\bsnm{Holz}, \binits{D.E.}},
\bauthor{\bsnm{Farr}, \binits{B.}}:
\batitle{Black hole coagulation: Modeling hierarchical mergers in black hole populations}.
\bjtitle{The Astrophysical Journal}
\bvolume{893}(\bissue{1}),
\bfpage{35}
(\byear{2020})
\end{barticle}
\endbibitem

\bibitem[\protect\citeauthoryear{Kimball et~al.}{2020}]{hierarchical}
\begin{barticle}
\bauthor{\bsnm{Kimball}, \binits{C.}},
\bauthor{\bsnm{Talbot}, \binits{C.}},
\bauthor{\bsnm{Berry}, \binits{C.P.L.}},
\bauthor{\bsnm{Carney}, \binits{M.}},
\bauthor{\bsnm{Zevin}, \binits{M.}},
\bauthor{\bsnm{Thrane}, \binits{E.}},
\bauthor{\bsnm{Kalogera}, \binits{V.}}:
\batitle{Black hole genealogy: Identifying hierarchical mergers with gravitational waves}.
\bjtitle{Astrophys. J.}
\bvolume{900},
\bfpage{177}
(\byear{2020})
\end{barticle}
\endbibitem

\bibitem[\protect\citeauthoryear{Kimball et~al.}{2021}]{Kimball:2020qyd}
\begin{barticle}
\bauthor{\bsnm{Kimball}, \binits{C.}}, \betal:
\batitle{{Evidence for Hierarchical Black Hole Mergers in the Second LIGO\textendash{}Virgo Gravitational Wave Catalog}}.
\bjtitle{Astrophys. J. Lett.}
\bvolume{915}(\bissue{2}),
\bfpage{35}
(\byear{2021})
\end{barticle}
\endbibitem

\bibitem[\protect\citeauthoryear{Tong et~al.}{2026}]{mass_gap_zenodo}
\begin{botherref}
\oauthor{\bsnm{Tong}, \binits{H.}}, et al.:
Evidence of the Pair Instability Gap in the Distribution of Black Hole Masses Data Release.
\doiurl{10.5281/zenodo.18222409}
\end{botherref}
\endbibitem

\bibitem[\protect\citeauthoryear{Collaboration et~al.}{2025}]{GWTC_4_PE_release}
\begin{botherref}
\oauthor{\bsnm{Collaboration}, \binits{L.S.}},
\oauthor{\bsnm{Collaboration}, \binits{V.}},
\oauthor{\bsnm{Collaboration}, \binits{K.}}:
GWTC-4.0: Parameter Estimation Data Release.
\doiurl{10.5281/zenodo.16053484}
\end{botherref}
\endbibitem

\bibitem[\protect\citeauthoryear{Collaboration and Collaboration}{2022}]{GWTC_2.1_PE_release}
\begin{botherref}
\oauthor{\bsnm{Collaboration}, \binits{L.S.}},
\oauthor{\bsnm{Collaboration}, \binits{V.}}:
GWTC-2.1: Deep Extended Catalog of Compact Binary Coalescences Observed by LIGO and Virgo During the First Half of the Third Observing Run - Parameter Estimation Data Release.
\doiurl{10.5281/zenodo.6513631}
\end{botherref}
\endbibitem

\bibitem[\protect\citeauthoryear{Collaboration et~al.}{2021}]{GWTC_3_PE_release}
\begin{botherref}
\oauthor{\bsnm{Collaboration}, \binits{L.S.}},
\oauthor{\bsnm{Collaboration}, \binits{V.}},
\oauthor{\bsnm{Collaboration}, \binits{K.}}:
GWTC-3: Compact Binary Coalescences Observed by LIGO and Virgo During the Second Part of the Third Observing Run — Parameter Estimation Data Release.
\doiurl{10.5281/zenodo.5546663}
\end{botherref}
\endbibitem

\bibitem[\protect\citeauthoryear{Collaboration et~al.}{2025}]{GWTC_4_sensitivity_release}
\begin{botherref}
\oauthor{\bsnm{Collaboration}, \binits{L.S.}},
\oauthor{\bsnm{Collaboration}, \binits{V.}},
\oauthor{\bsnm{Collaboration}, \binits{K.}}:
GWTC-4.0 Cumulative Search Sensitivity Estimates.
\doiurl{10.5281/zenodo.16740128}
\end{botherref}
\endbibitem

\end{thebibliography}
\end{document}